\documentclass[acmsmall]{article}

\usepackage{booktabs}
\usepackage{url}
\usepackage{amsmath}
\usepackage{subcaption}
\usepackage{todonotes}
\usepackage{enumitem}
\usepackage{listings}
\usepackage{minted}
\setminted[c]{
    fontsize=\footnotesize,
    numbersep=-8pt,
    obeytabs=true,
    linenos
}
\usepackage[numbers]{natbib}

\newcommand{\acmcaption}[1]{\caption{#1}}

\usepackage{algorithm}
\usepackage{algorithmic}

\usepackage[dvipsnames]{xcolor}
\definecolor{torvergreen}{RGB}{000, 125, 52}
\lstdefinestyle{algo}{
  belowcaptionskip=1\baselineskip,
  breaklines=true,
  frame=lines,
  xleftmargin=\parindent,
  showstringspaces=false,
  basicstyle=\footnotesize\fontfamily{cmr}\selectfont,
  keywordstyle=\bfseries\color{green!40!black}\fontfamily{cmtt}\selectfont,
  commentstyle=\itshape\color{purple!40!black},
  emph={for, do, end},
  emphstyle=\color{red},
  keywords={MPI_Comm_size, fun, MPI_Isend, MPI_Iprobe, MPI_Get_count, MPI_Recv},
  stringstyle=\color{orange},
}

\usepackage{soul}
\usepackage[margin=2.5cm]{geometry}

\newcommand{\best}[1]{\color{torvergreen}{\underline{#1}}}

\begin{document}

\title{When High-Performance Computing Meets Software Testing: Distributed Fuzzing using MPI}

\author{Pierciro Caliandro, Matteo Ciccaglione, Alessandro Pellegrini}
\maketitle

\begin{abstract}
This paper explores the integration of MPI-based synchronization techniques into distributed fuzzing frameworks, highlighting possible substantial performance improvements compared to traditional filesystem-based synchronization methods. By employing lightweight MPI primitives, reductions in communication latency are achieved, facilitating more efficient data exchanges across distributed fuzzing nodes. Experimental results obtained over standard benchmarks demonstrate enhanced coverage progression from the early stages of the fuzzing process, which could be beneficial if fuzzing is employed in CI/CD pipelines at any stage of software development. Furthermore, the coordinated exchange of input corpora among clusters of fuzzers effectively addresses coverage stagnation, enabling a sustained exploration of complex and deep execution paths. Overall, the adoption of MPI-based synchronization approaches shows promising potential for significantly enhancing the scalability and efficacy of distributed fuzz testing.
\end{abstract}

\section{Introduction}

Fuzzing is an automated software testing technique that systematically generates and injects random or semi-random inputs into a program to uncover vulnerabilities, errors, or unexpected behaviour~\cite{Li2018-np}. By rapidly bombarding software interfaces with diverse and unconventional data patterns, fuzzing aims to trigger conditions that developers might overlook during manual testing. This method proves particularly effective at discovering critical security flaws such as buffer overflows, memory corruption, or unhandled exceptions, making it indispensable in enhancing software robustness and reliability. Its utility spans various domains from simple command-line utilities to complex network protocols, highlighting fuzzing's pivotal role in contemporary cybersecurity practices.

Driving a successful fuzzing campaign involves implementing targeted strategies that leverage advanced techniques such as coverage-based fuzzing, mutation-based fuzzing, and generation-based fuzzing. Coverage-based fuzzing utilises feedback from the program's execution to identify unexplored code paths, thus guiding the input generation to maximise code coverage and discover hidden vulnerabilities more effectively. Mutation-based fuzzing modifies existing input samples incrementally, observing how slight variations influence program behaviour and stability. Generation-based fuzzing, on the other hand, systematically constructs inputs based on defined models or specifications, aiming to cover scenarios that random mutation might miss. Key to these methods is a robust testing environment capable of logging inputs, outputs, and any abnormal behaviour. Continuous monitoring, along with result analysis, provides actionable insights that enable iterative enhancement of test cases and refinement of fuzzing approaches, significantly improving the detection of subtle yet critical software issues.

The time required for fuzzing an application varies significantly, depending on factors such as software complexity, input space size, and desired coverage depth. Comprehensive fuzzing campaigns may extend from several hours to weeks or even months, especially for complex systems with intricate internal API or large software libraries. While prolonged fuzzing can uncover elusive vulnerabilities that surface only under rare conditions, thus significantly enhancing software security, it can also be a notable impediment. Extensive fuzzing demands substantial computational resources and incurs high costs, potentially delaying software release cycles. Therefore, balancing thoroughness and practicality is essential to optimise fuzzing effectiveness without adversely impacting the development timeline or budget.

While modern fuzzers can already be executed in parallel by simply launching multiple independent processes on the same machine, this naïve strategy is fundamentally limited. Such uncoordinated parallelism may result in non-negligible redundancy: different fuzzing instances may waste cycles evaluating identical inputs or become independently stuck on the same hard-to-fuzz code paths. Meanwhile, one instance might discover a useful input that could help others progress. Without coordination, this knowledge is not shared.

The lack of communication between fuzzing processes significantly reduces the effective throughput of the campaign, diminishing the advantages of parallel execution. To mitigate this, advanced fuzzing systems employ inter-process communication to synchronise input corpora and share coverage feedback. Traditional solutions rely on the file system, which introduces latency, while more efficient approaches use shared memory abstractions such as \texttt{shm\_open()} or \texttt{mmap()} with \texttt{MAP\_SHARED}, as available on Linux. This shift reflects a broader evolution: fuzzing has matured into a high-performance computing workload, as exemplified by industrial-scale efforts like Google OSS-Fuzz \cite{Serebryany2017-nq}, and must be engineered accordingly to overcome the stagnation of single-threaded CPU scaling~\cite{Ross2008-fb}.

In this paper, we propose a methodology designed to enhance coverage-based distributed fuzzing by leveraging the Message Passing Interface (MPI)~\cite{Message-Passing-Interface-Forum2023-ou} for parallel computation, employing techniques that have long been effectively utilised in the High-Performance Computing (HPC) community. MPI, widely adopted for large-scale parallel scientific computations such as simulations in computational fluid dynamics~\cite{Afzal2017-ta}, parallel discrete event simulation~\cite{Carothers2002-ue,Martin2003-rh, Pellegrini2012-eo}, molecular dynamics (e.g., GROMACS~\cite{Van-Der-Spoel2005-tk}, LAMMPS~\cite{Thompson2022-el}), and weather modelling (e.g., WRF~\cite{Christidis2015-gz}), provides robust mechanisms for efficient inter-process communication and coordination. Our approach introduces several input distribution strategies among MPI ranks, including \textit{static partitioning}, \textit{dynamic load balancing}, and \textit{feedback-driven workload assignment}. Static partitioning assigns predefined input subsets to specific MPI ranks, facilitating predictable coverage distribution. Dynamic load balancing redistributes fuzzing workloads among ranks based on runtime metrics such as processing speed or input complexity, optimising computational resource utilisation. Feedback-driven workload assignment integrates runtime coverage feedback to dynamically adjust input distribution, enhancing the efficiency of exploring novel code paths. Empirical evaluations across various application classes demonstrate that our methodology can effectively reduce the time required to achieve targeted coverage goals, consequently accelerating the discovery of software vulnerabilities and defects.

We have integrated a reference implementation of our proposals in AFL++ fuzzing framework~\cite{Fioraldi2020-pz}. AFL++ is notable for its technical capabilities, particularly its instrumentation-based feedback mechanism, which enables accurate monitoring of executed code paths. 
We compare our proposed MPI-based distributed fuzzing methodology against two alternative strategies: a naïve tar-based distributed fuzzing approach and an advanced method utilising the off-the-shelf Network File System (NFS) distributed filesystem. The naïve tar-based strategy involves periodically compressing and transferring input corpora across fuzzing nodes, which, while straightforward, suffers from inefficiencies related to frequent data synchronisation and overhead from repeated compression and decompression cycles. Conversely, NFS-based fuzzing leverages optimised distributed storage mechanisms, theoretically offering improved synchronisation and reduced overhead. 

Our evaluation examines metrics such as synchronisation latency, coverage progression, and the overall efficacy in detecting software vulnerabilities. We rely on a wide set of standard benchmark applications to assess the benefits of our proposal. In particular, we identify specific scenarios where each approach excels or falls short. Our results demonstrate that the MPI-driven strategies provide significant improvements in performance, efficiency, and scalability, particularly in large-scale distributed fuzzing scenarios, by effectively overcoming the synchronisation bottlenecks and redundancy challenges inherent to alternative solutions.

\section{Related Work}
\label{sec:related}

The literature on parallel fuzzing is not particularly extensive, with a few exceptions in recent years. The reason for this is mainly linked to the idea that, given enough time, even local fuzzing is capable of identifying a sufficient number of defects. However, there are various studies showing that, by effectively exploiting a distributed architecture, it is possible to achieve good results in less time, thus allowing fuzzing campaigns to be implemented even in contexts where completion time is more important, such as in CI/CD pipelines during incremental development activities.

Some works~\cite{Pham2021-zv,Wang2021-zj,Luo2024-qj} introduce fuzzing frameworks that exploit program structures, such as call graphs or control-flow graphs, to systematically allocate fuzzing tasks across parallel instances. They focus on macro-task partitioning based on program semantics, aiming to reduce redundancy and enhance fuzzing efficiency. In particular, the works in~\cite{Pham2021-zv,Luo2024-qj} propose a graph-partitioning approach to dynamically allocate macro-tasks based on attributed call graphs. The frameworks integrate dynamic fuzzing data to iteratively refine task allocation, significantly improving branch coverage and bug detection efficiency. Conversely, in~\cite{Wang2021-zj}, the authors explicitly tackle the reduction of task overlap through seed-centric, dynamic assignment of tasks to fuzzing instances. Some of our policies have a similar rationale, although we explicitly take into account the possibility that this clustering can be stuck in the exploration of the fuzzed applications, and we propose a countermeasure to such scenario. In this sense, differently from~\cite{Wang2021-zj}, we explicitly allow some overlap as it may be beneficial to escape stall situations. Additionally, we explicitly account for different fuzzing strategies in some of our policies.

Centralized databases or data storage mechanisms to synchronize fuzzing states among distributed fuzzing instances have been explored in~\cite{Song2019-da,Zhou2023-ps}. Particularly, in~\cite{Song2019-da} the authors rely on MongoDB to synchronize seeds and coverage bitmaps, while in~\cite{Zhou2023-ps} a dedicated centralized server implemented using Berkeley sockets is used for a similar goal. We do not implement any centralized server, as we are interested in exchanging the input corpora between the fuzzers according to different policies. In addition, the work in~\cite{Zhou2023-ps} proposes a hierarchical synchronization to reduce resource utilization. We share a similar strategy, but with the goal of handling different fuzzing approaches in a coordinated way.

The work in~\cite{Chen2018-ph} proposes an ensemble fuzzing approach integrating multiple diverse fuzzers (AFL variants, libFuzzer, Radamsa, QSYM) with a seed synchronization mechanism (GALS), enabling effective collaboration and higher robustness in bug discovery. Having different fuzzers is also a shared goal with some policies that we propose in this paper. Overall, our proposal aims at exploiting parallelism and optimizing computational resource use at large scale through MPI, with a significant focus on inter-process communication efficiency and adaptive workload distribution. Conversely, the authors of~\cite{Chen2018-ph} mainly focus on strategic diversity among fuzzers to increase robustness against target applications by exploiting different fuzzing heuristics. In this perspective, the two proposals can benefit from each other.

Some works~\cite{Li2018-lk,Zhou2023-ps} exploit distributed fuzzing as a way to reduce synchronization overhead and improve timely exploration, as we do. In~\cite{Li2018-lk}, the authors address synchronization overhead via simplification of execution paths into compact, linear array representations. This strategy allows to efficiently manage synchronization among nodes. Similarly, in~\cite{Zhou2023-ps}, the authors propose a centralized, dynamic scheduling approach that optimizes resource allocation by dispatching fuzzing tasks and adaptively balancing workloads. Our work is more general, because we propose a synchronization infrastructure based on MPI that can be integrated with differentiated policies to exchange the input corpora between fuzzers.

A different level of distributed fuzzing is showcased in~\cite{Jang2022-lj}. Here, the authors present a highly distributed infrastructure, named Fuzzing@Home, which targets large-scale, heterogeneous, untrusted public computing resources. Our reference infrastructure is more proper of HPC facilities, like supercomputers, operating on a local network. Therefore, in~\cite{Jang2022-lj}, the authors are more interested in the trustworthiness of the network, with special focus on identifiying possible malicious actors. At the same time, this work confirms the validity of relying on distributed fuzzing as a way to enhance the possibility of discovering new bugs thanks to an increased amount of computing power. Therefore, the two works are inherently orthogonal, but increase each other's significance.

\section{Distributed Fuzzing with MPI}

In this Section, we present and discuss various synchronization techniques tailored for distributed fuzzing using MPI. These policies vary significantly in complexity and underlying assumptions, enabling us to explore their effectiveness across diverse fuzzing contexts. With these policies, we aim to identify mechanisms that best exploit computational resources and adapt to the specific demands posed by different applications.

The principal goal of these policies is to enhance fuzzing efficiency, specifically by reducing the time required to reach some target coverage of the fuzzed application. Given that improved coverage directly correlates with a higher likelihood of discovering latent software bugs, these policies can effectively accelerate the fuzzing process, increasing vulnerability detection rates.

\subsection{Selective Dissemination Policy}

The first simple policy that we introduce to optimize synchronization in our distributed fuzzing approach, employs a selective dissemination mechanism based on hashing. The key rationale is that when an individual fuzzing instance identifies a valuable input---typically an input that enhances code coverage---it should propagate this discovery only to specific peer instances that might benefit from it, rather than broadcasting it universally. This selective approach avoids redundant processing and unnecessary synchronization overhead, thereby improving resource utilization and overall performance.

Our strategy employs a hash-based function to determine the destination instance for each valuable test case. Specifically, upon identifying a beneficial input, the originating fuzzing node calculates a hash of this input using the 64-bit variant of the widely used \texttt{xxhash} function\cite{ColletUnknown-rt}. The target fuzzing instance is then selected based on the following modulo-based mapping:

\begin{equation*}
    \text{instance\_rank} = H(\text{test\_case}) \mod N
\end{equation*}

\noindent
where $H$ represents the 64-bit hash value computed with \texttt{xxhash}, and $N$ corresponds to the total number of instances within the MPI communicator (\texttt{MPI\_Comm\_size}). This deterministic approach ensures a uniform distribution of test cases across all fuzzing nodes, balancing the load effectively. The uniformity and low collision rate of the chosen hash function ensure minimal synchronization overhead and equitable sharing of computational workloads among the nodes.

This policy has the merit of significantly reducing the redundancy common in naive parallel fuzzing setups, wherein valuable inputs are either disseminated too broadly, causing unnecessary duplication of effort, or not disseminated efficiently, reducing potential coverage gains. At the same time, the obvious drawback of this policy is that the deterministic selection of the target of the input may not pick the fuzzing instance that could benefit most from adding it to their corpora, resulting in increased fuzzing time with no actual gain. Moreover, although fast, computing a hash on the input with \texttt{xxhash} adds some computation overhead on the critical path of the fuzzer, which could delay the execution of relevant test cases and their dissemination.

\begin{figure}[!ht]
	\includegraphics[width=\linewidth]{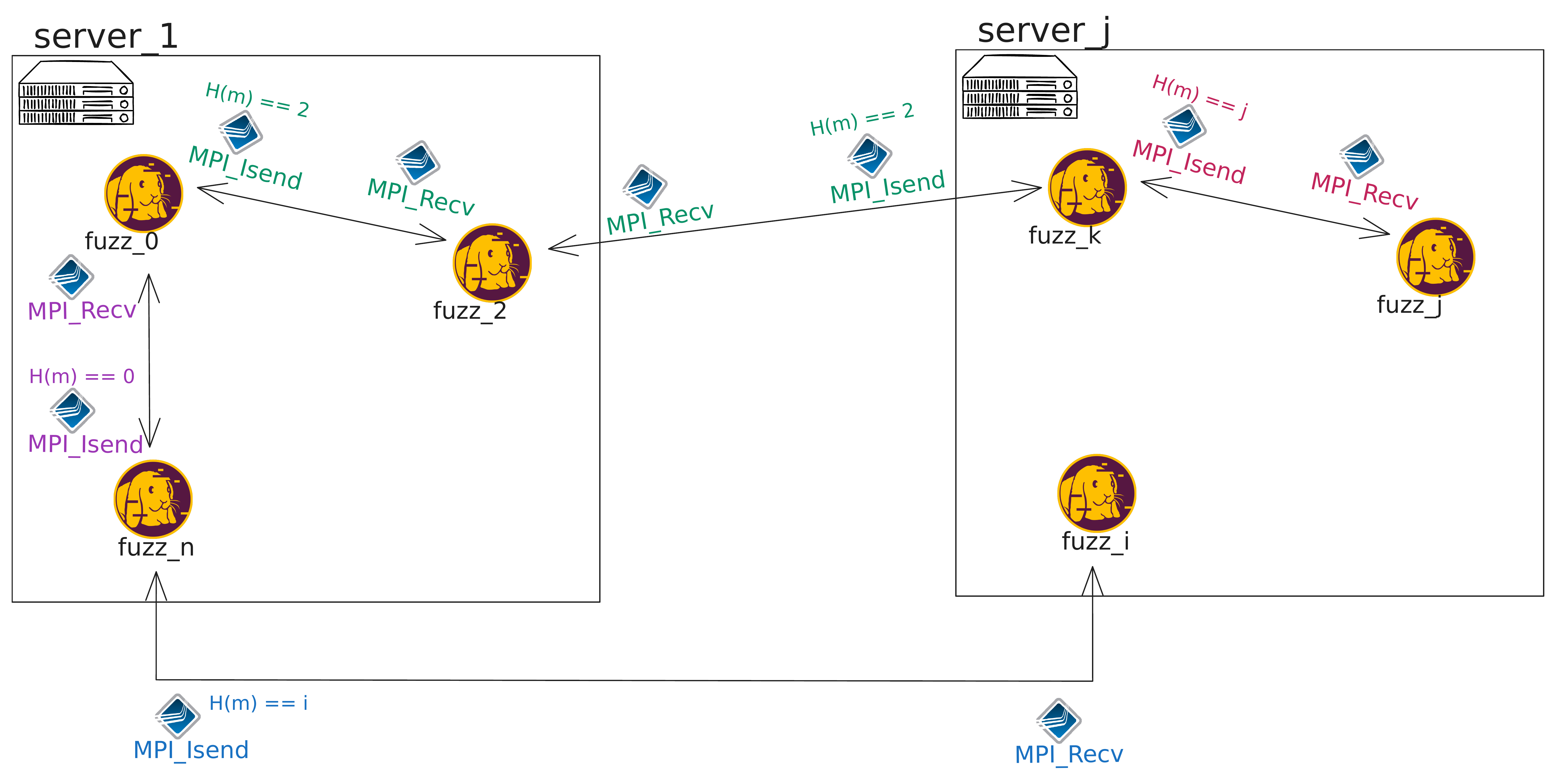}
	\acmcaption{Selective Strategy}
	\label{fig:selective-strategy}
\end{figure}

\subsection{Dynamic Dissemination with Utilization}
\label{sec:dynamic-dissemination}

To solve some of the drawbacks of the previous baseline policy, we introduce a slightly more sophisticated dissemination policy named \textit{Dynamic Dissemination with Utilization Feedback}. Unlike the previous, this strategy employs a feedback mechanism that dynamically adjusts input dissemination based on actual utilization and effectiveness observed at the receiving instances.

In this policy, each fuzzing node maintains a dedicated data structure called the \textit{utility directory}. The utility directory acts as a dynamic registry that tracks the historical effectiveness of inputs received from other instances, associating sender ranks with corresponding utilization metrics. Concretely, the directory is implemented as a hash table mapping sender ranks to a utilization score, incremented each time an incoming test case from a sender contributes positively to code coverage, and decremented otherwise.

When an instance generates a new test case that it deems valuable, it first consults the utility directory to identify the instance historically demonstrating the highest utilization rate of previously sent inputs. Thus, rather than selecting a static dissemination target as in the hash-based policy, dissemination dynamically adapts, prioritizing recipients based on measured effectiveness. Upon receiving an input, a node derives its value according to standard criteria (e.g., improved coverage). If the input is deemed beneficial, the receiver increments the sender’s score in its local utilization directory. Conversely, if the input is repeatedly unproductive (thus consuming computational resources without benefit), the receiver decreases the corresponding score. If this score drops below a predefined threshold, the receiver transmits a control message back to the sender, indicating that inputs from that sender have low utility, asking to stop sending inputs. Upon receiving such feedback, the sender removes the receiver from the set of interested ranks.

This feedback-driven approach can improve the adaptive dissemination of test cases by ensuring that computational resources are allocated toward beneficial fuzzing activities. However, in scenarios characterized by highly dynamic code coverage evolution, rapid oscillations in dissemination preference could lead to an unnecessary partitioning of the nodes, possibly preventing relevant inputs to be received in subsequent phases of the fuzzing campaign.

\begin{figure}[!ht]
	\includegraphics[width=\linewidth]{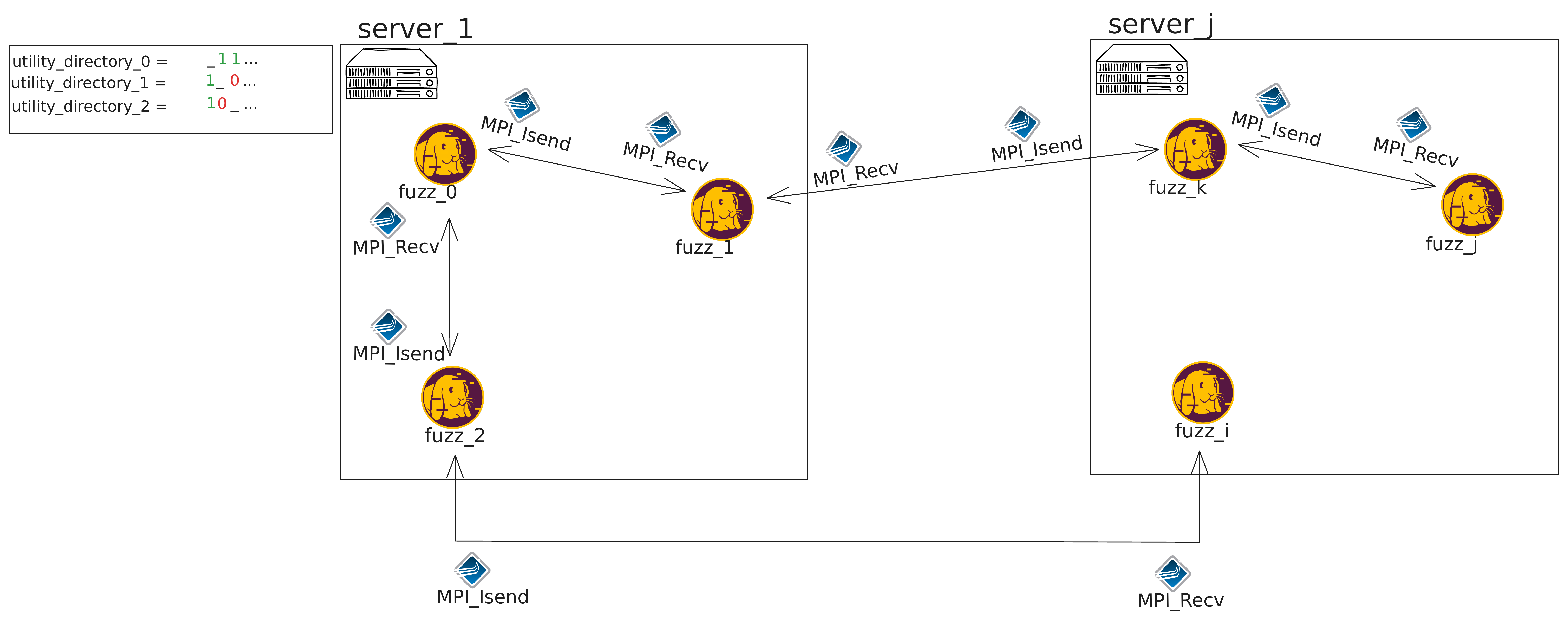}
	\acmcaption{Dynamic strategy overview}
	\label{fig:dynamic-strategy}
\end{figure}

\subsection{Hierarchical Dissemination Across Fuzzer Clusters}

The aforementioned policies do not take into account a fundamental aspect related to fuzzing. Typically, fuzzing campaigns rely on different fuzzing instances that employ different instrumentation methods and mutation strategies---such as AddressSanitizer (ASAN), LAF
instrumentation, comparison logging (cmplog), or power-scheduling strategies---that yield distinct exploration patterns and code coverage profiles. Indiscriminately exchanging inputs across fuzzers employing radically different approaches could be inefficient, since inputs valuable to one strategy may not be equally beneficial to another, potentially causing a waste of resources.

The third policy that we consider in this work structures the distributed fuzzing environment into logical clusters, each consisting of fuzzing instances employing homogeneous or closely related instrumentation and mutation strategies. Within each cluster, fuzzing nodes adhere to a straightforward synchronization pattern: secondary nodes disseminate discovered inputs exclusively to a designated master node, forming a simple tree-based hierarchy.

Each master node maintains a cluster-specific input corpus, continuously aggregating inputs discovered by its secondary instances. To enable controlled inter-cluster collaboration, these master nodes periodically synchronize their corpora with peer masters from other clusters. This strategy can be integrated with the one described in Section~\ref{sec:dynamic-dissemination}, enabling a selective cross-cluster synchronization based on utility metrics, such as incremental coverage gain, thus preventing unnecessary propagation of unproductive inputs.

The hierarchical clustering strategy offers substantial advantages: it reduces synchronization overhead by constraining frequent input exchanges within clusters sharing common fuzzing characteristics (thus promoting specialization), enhances input relevance by selectively cross-synchronizing clusters according to observed utility, and scales effectively in highly distributed setups by limiting communication overhead. Overall, this strategy can maximise coverage effectiveness through specialised mutation strategies, while still benefiting from occasional high-quality inputs derived from complementary fuzzing methodologies.

\begin{figure}[!ht]
	\includegraphics[width=\linewidth]{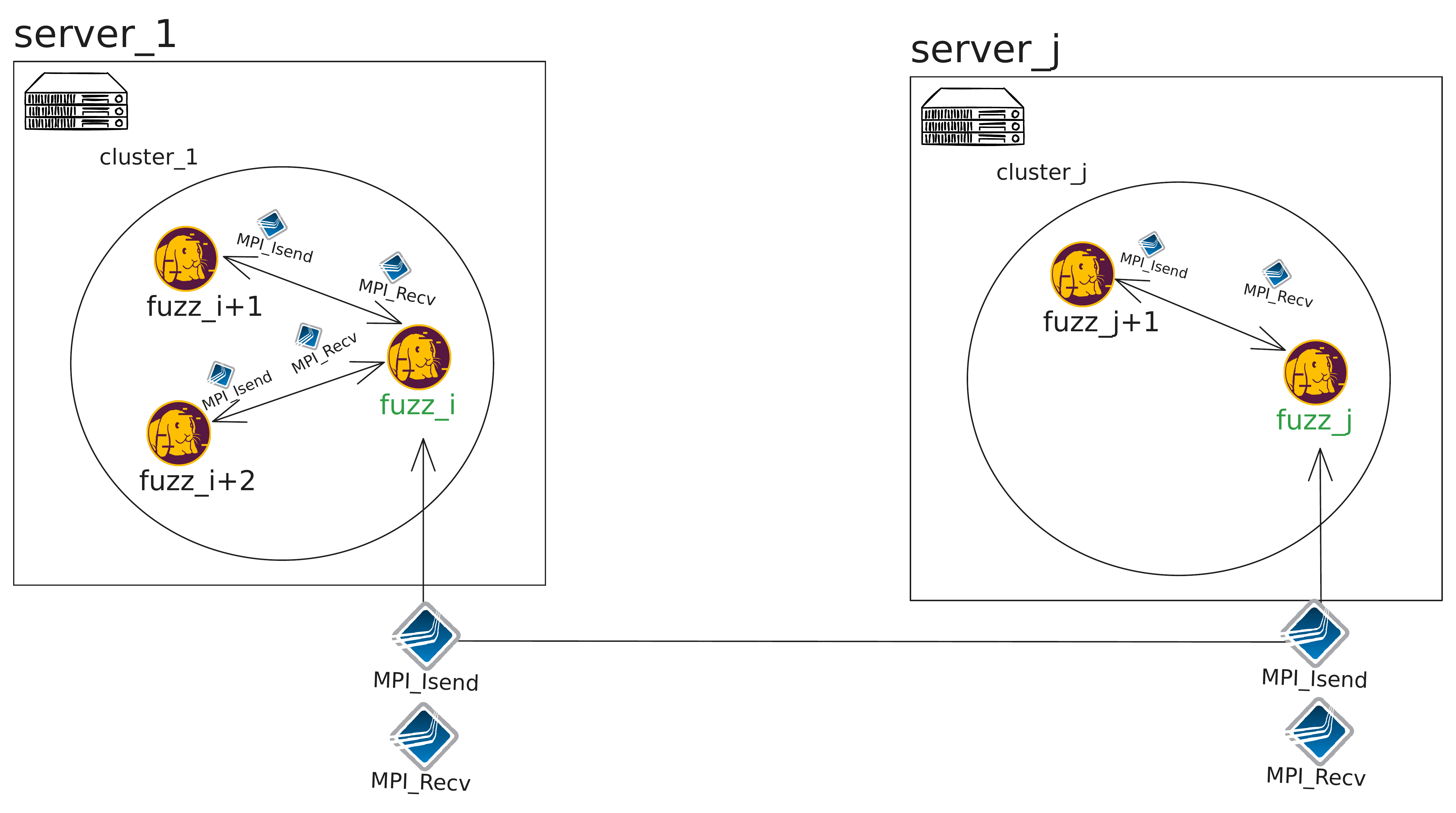}
	\acmcaption{Hierarchical Dissemination Strategy}
	\label{fig:hierarchical-strategy}
\end{figure}

\subsection{Ammuina Mode}

All the previously discussed strategies can occasionally experience prolonged coverage stagnation, which may be observed if the fuzzing campaign is dealing with complex or deeply nested code paths.  A drastic change in the input corpora may allow fuzzers to get away fast from this stagnation. Therefore, we introduce an adaptive synchronization mode named \textit{ammuina mode}%
\footnote{The name stems from the phrase ``\textit{facite ammuina}'', a humorous Neapolitan expression, roughly translating to ``make confusion'' or ``create chaos.'' Popularly (and erroneously) believed to originate from a fabricated 19th-century Bourbon navy command instructing sailors to feign activity and disorder aboard ships to deceive observers, the term endures in contemporary Italian as a playful critique of superficial busyness or intentional distraction without real productivity.}. This trigger-based synchronization mechanism explicitly targets scenarios where fuzzing progress stalls, complementing and enhancing the previously discussed dissemination policies.

The fundamental principle behind ammuina mode is to periodically detect when individual fuzzing nodes reach a stagnation point, characterized by negligible incremental coverage gains. Specifically, this detection is performed by each fuzzing instance when periodically updating its internal statistics. The instances evaluate two conditions: (i) whether their current coverage increment is below a predefined threshold $t_{inc}$, and (ii) whether sufficient time, defined by the threshold $t_{time}$, has elapsed since the last time the coverage increment was above $t_{inc}$. If both conditions hold, the instance broadcasts a request to initiate ammuina mode synchronization.

Consequently, all nodes transition into ammuina mode, orchestrating an input exchange aimed at revitalizing coverage progress. The nodes first disseminate locally-discovered inputs deemed potentially valuable, and receive inputs from the other ones, using asynchronous MPI scatter\slash gather primitives. In this way, all nodes incorporate externally-generated inputs into their local corpora, in the attempt to overcome previously challenging coverage barriers.

Clearly, the drawback of this approach is is the possible increased communication overhead, potentially causing temporary slowdowns across the distributed fuzzing campaign. Moreover, the values of the thresholds could play a significant role: overly aggressive synchronization could result in frequent, unnecessary, and unfruitful slowdowns, while excessively conservative thresholds might limit the effectiveness of this approach.

\subsection{Highlights on the Reference Implementation}

Our proposal builds upon and integrates with AFL++, a widely adopted framework for automated software testing through fuzzing. AFL++ supports the concurrent execution of multiple fuzzing instances, either on the same target binary or across different ones. In particular, it provides native support for parallel, though not distributed, fuzzing campaigns. When executed on a multi-core system, AFL++ employs a primary\slash secondary architectural pattern in which a designated *main node* manages the shared input corpus and coordinates a number of *secondary nodes*, each of which is bound to a separate CPU core and operates independently.

AFL++ incorporates advanced mutation strategies that systematically vary inputs, effectively increasing the probability of triggering unforeseen program states. It also integrates complementary methodologies such as symbolic execution and concolic testing, enhancing the exploration of code paths that are difficult to reach through random input generation alone. These technical features collectively enable AFL++ to provide comprehensive and thorough fuzz testing, making it a widely utilised tool within academic research and practical software security assessments.

AFL++ suffers from the same time constraints mentioned previously, primarily due to the inherently exhaustive nature of fuzz testing, which involves extensive exploration of execution paths and mutation of inputs. Distributed fuzzing presents a viable approach to mitigate these time constraints by parallelising the fuzzing process across multiple machines or CPU cores, effectively speeding up coverage exploration. However, AFL++ currently offers limited built-in support for effective distributed fuzzing. Although it includes basic parallel fuzzing capabilities, such as running multiple instances concurrently, it lacks advanced synchronisation and load balancing mechanisms necessary to maximise resource utilisation and avoid redundancy. Therefore, to achieve effective distributed fuzzing with AFL++, external tools or custom scripts are often required to manage multiple instances efficiently, track shared code coverage, and systematically coordinate the fuzzing process.

In the reference implementation we embed MPI primitives directly inside AFL++ so that inter-node communication no longer depends on the operating-system file system. All fuzzing processes are enrolled in the default \texttt{MPI\_Comm\_world}; consequently every instance can communicate with any other via the standard rank abstraction. The first point of intervention is the routine \texttt{save\_if\_interesting} of AFL++. Whenever AFL++ decides that a test case has sufficient novelty, the corresponding input buffer is now dispatched with a non-blocking \texttt{MPI\_Isend} rather than being written to disk, using the relevant scheme adopted by the different policies. Because the sender must later reclaim the buffer, every outstanding transmission is recorded in a lightweight queue whose nodes hold the pointer to the message and the associated \texttt{MPI\_Request}.

\begin{figure}[!ht]
	\includegraphics[width=0.7\linewidth]{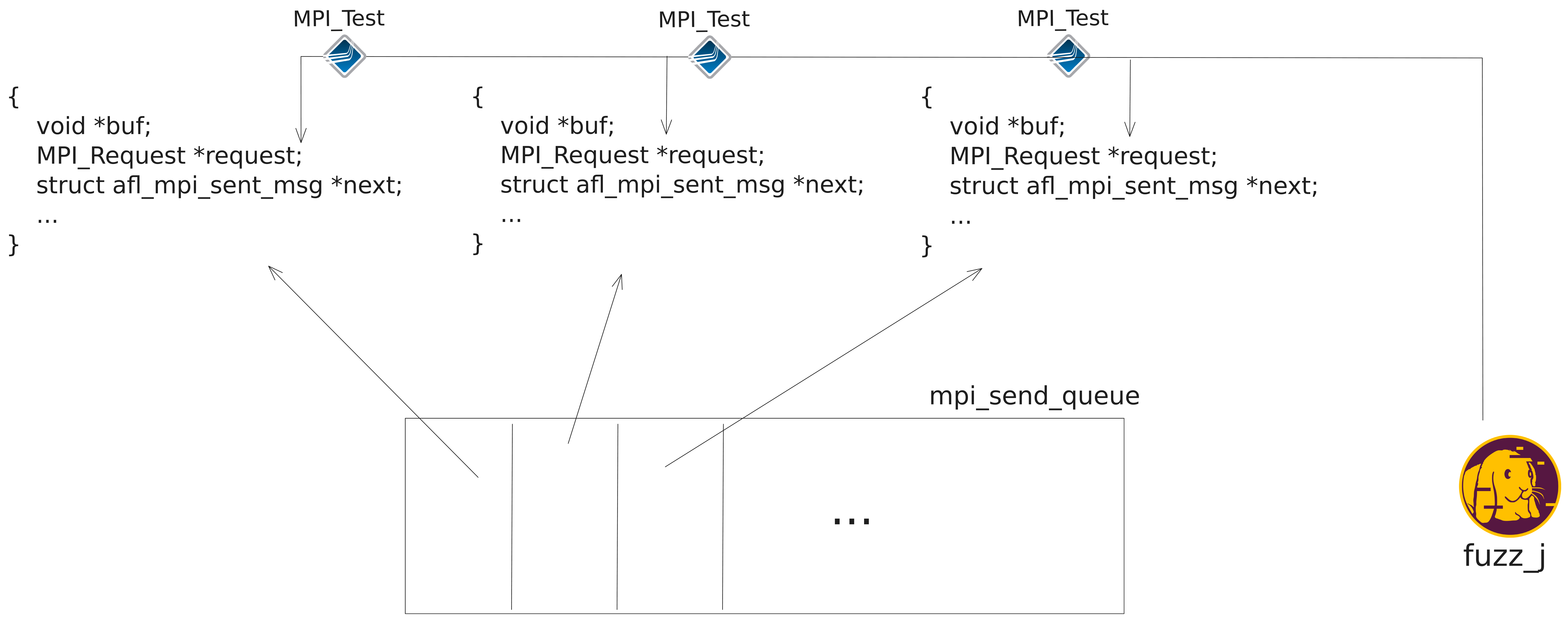}
	\acmcaption{MPI message send queue}
	\label{fig:mpi_send_queue}
\end{figure}

The queue, illustrated in Figure~\ref{fig:mpi_send_queue}, is a global structure local to each rank and is periodically scanned: if \texttt{MPI\_Test} reports completion, the request is removed and the buffer is released, ensuring that the asynchronous path never exhausts memory. 

Reception is handled directly in \texttt{sync\_fuzzers}: here, \texttt{MPI\_Iprobe} is used to check for pending messages from \texttt{MPI\_ANY\_SOURCE} and, in the positive case, the incoming message is extracted with  \texttt{MPI\_Recv}. Once a buffer arrives, the contained input is tested exactly as in ALF++, calling \texttt{fuzz\_run\_target} and subsequently invoking \texttt{save\_if\_interesting} to verify if such input contributed to novel code path discovery.


\section{Experimental Assessment}
\label{sec:experimental}

In this Section, we present the experimental data that we have collected to study the behaviour of the different MPI-based fuzzing policies. We have used standard benchmarks from the literature to study the fuzzing behaviour, and we have compared our proposal with different baseline configurations. All these aspects are detailed below.

\subsection{Selected Benchmarks and Testbed Configuration}
\label{sec:benchmarks}

To stress the policies that we have introduced in this article, we have selected 10 different applications from Google OSS-Fuzz~\cite{Serebryany2017-nq} benchmark suite, namely:

\begin{itemize}[noitemsep]
    \item \textit{FreeType2}: a popular font rendering engine enabling high-quality text rasterization, which requires complex data inputs;
    \item \textit{Guetzli}: a JPEG encoder developed by Google, that optimizes image compression to generate high-quality JPEG images with significantly smaller sizes compared to standard encoders;:
    \item \textit{HarfBuzz}: a text-shaping engine, which supports complex text layouts and scripts, useful in rendering multilingual typography and used extensively in graphical applications and browsers;
    \item \textit{LCMS} (little CMS): a compact colour management engine, widely employed in software that requires accurate color transformation and profile handling;
    \item \textit{Libjpeg}: a widely-used library to encode\slash decode JPEG images, often used as a performance and correctness benchmark;
    \item \textit{Libpng}: the official reference library to manipulate PNG images;
    \item \textit{PCRE} (Perl Compatible Regular Expressions): a commonly used library providing functionality for pattern matching and to parse text using regular expressions; 
    \item \textit{PROJ}: a generic coordinate transformation software, that transforms coordinates from one coordinate reference system (CRS) to another;
    \item \textit{RE2}: Google's regular expression library, developed with a focus on performance;
    \item \textit{WOFF2}: a library implementing the Web Open Font Format 2 compression standard, used to reduce font file size for web delivery;
\end{itemize}

All these applications and libraries are well-known open source libraries, that are commonly used in fuzzing campaigns to benchmark the effectiveness of different fuzzing strategies.

We have run our experiments on a cluster composed of 3 machines, each one equipped with two Intel(R) Xeon(R) E5-2630 v3@2.40GHz CPUs and 128 GB of DDR4 RAM, for a total of 111 CPU cores. We have used OpenMPI~\cite{Gabriel2004-nr} as the underlying MPI communication library.

\subsection{Baseline AFL++ Configurations}

All fuzzing instances have been configured adhering to the official guidelines and best practices provided by the AFL++ developers for parallel fuzzing campaigns. These practices are documented in detail in the AFL++ technical documentation%
\footnote{\url{https://aflplus.plus/docs/fuzzing_in_depth/} [Last Accessed on June 2, 2025].},
which we closely followed to configure and execute our experiments.

A critical aspect concerns the diversification of fuzzing strategies across instances. Specifically, when employing multiple fuzzers in parallel, it is advised to compile multiple versions of the target binary, each with different instrumentation flags, to maximise coverage and bug discovery potential. At a minimum, one instance should be configured to fuzz a binary compiled with comprehensive memory sanitizers enabled, including AddressSanitizer, UndefinedBehaviorSanitizer, and Control Flow Integrity Sanitizer (achieved by setting \texttt{AFL\_USE\_ASAN=1}, \texttt{AFL\_USE\_UBSAN=1}, and \texttt{AFL\_USE\_CFISAN=1}, respectively). Additionally, one or two instances should employ the \textit{redqueen mode}~\cite{Aschermann2019-kr}, with at least one of them enabling input-to-state transformations to enhance path sensitivity. A further subset of instances should target binaries instrumented with the \texttt{laf-intel} or \texttt{COMPCOV} enhancements, which are designed to make complex control-flow structures more transparent to the fuzzer. Any remaining secondary nodes should be allocated to fuzzers configured with a heterogeneous mix of AFL++ strategies, thereby increasing the diversity of mutation and exploration heuristics in the campaign.

This configuration paradigm reflects a strategy of \textit{instrumentation heterogeneity}, where the combined effect of distinct fuzzing heuristics and compiler instrumentation aims to cover a broader space of program behaviours. It is also aligned with empirical findings suggesting that fuzzing effectiveness is significantly improved when input diversification and instrumentation diversity are maximized~\cite{Gan2018-tn}. We have retained this heterogeneity in all fuzzer configurations, also when employing our MPI-based synchronization policies.

\begin{figure}
    \centering
    \includegraphics[width=0.8\linewidth]{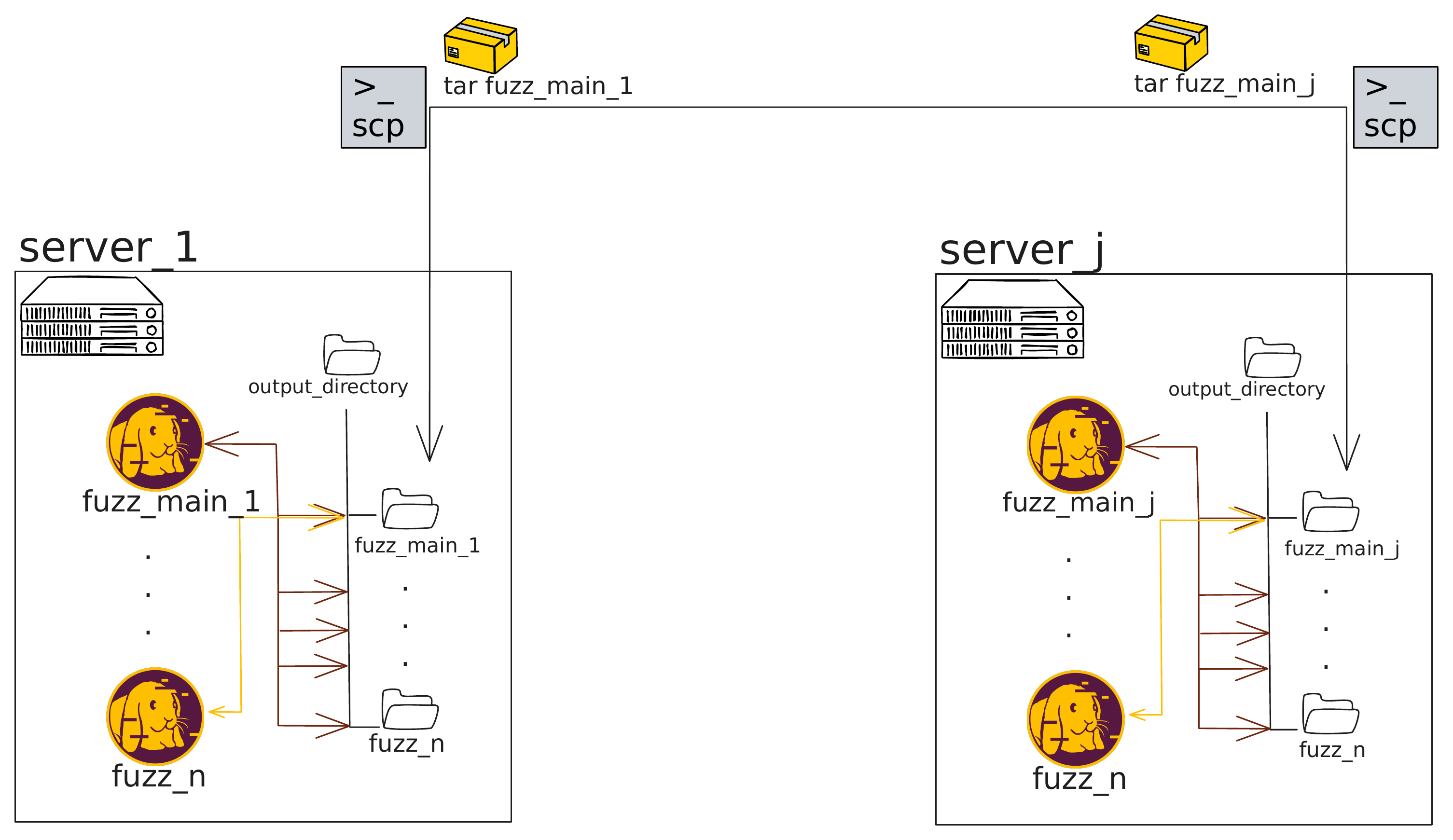}
    \acmcaption{\texttt{AFLplusplus sync strategy}}
    \label{fig:app_sync}
\end{figure}

We have used two baseline configurations based on the AFL++ fuzzer. The first baseline, which we refer to as \textsf{afl++} in the results, is a multi-node distributed configuration of AFL++ which is based on explicit, periodic synchronization of the input corpora. This baseline is also based on the AFL++ official documentation. The various distributed nodes grab interesting inputs from the others as depicted in Figure~\ref{fig:app_sync}: every secondary node grabs input from the main node while the main node syncs itself with any other node. The data exchange between main nodes follows once again the guideline written in the technical documentation%
\footnote{\url{https://aflplus.plus/docs/fuzzing_in_depth/\#d-using-multiple-machines-for-fuzzing}}, employing a shell script that periodically compresses the queue directory of each main node and transfers it to the synchronization directory of the others.

The second baseline configuration is based on a distributed network file system. In particular, we have configured parallel instances of AFL++ in the various nodes, which synchronize their input corpora using standard file system primitives, but the data are kept in a shared mountpoint based on NFS. This configuration, named \textsf{nfs} in the results, allows us to study whether standard practices for file exchange on a local network are competitive compared to our MPI-based synchronization policies.

By default, if the instances are launched on the same machine, the underlying file system is used, where each fuzzer has its own subdirectory under the path \texttt{/path/to/sync\_dir/}.

\subsection{Results}

The benchmarks discussed in Section~\ref{sec:benchmarks} have undergone a long-lasting fuzzing campaign using all the baseline and MPI-based policies. We have run each fuzzing campaign for 6 hours. In the results, we provide the data only for the first hour of fuzzing, because we are mostly interested in what happens at the beginning of the fuzzing campaing: as mentioned, we are interested in understanding whether distributed fuzzing can play a relevant role when we want to be fast at exploring the application behaviour, e.g. for the integration of fuzzing in frequently-executed CI/CD pipelines.

Each MPI-based policy has been assessed with and without the special ammuina mode. When this mode is active, we report in the plots a vertical dashed line every time that the MPI ranks agreed upon starting an ammuina exchange---vertical lines are colored with the same color of the associated policy. For each benchmark, we report the total branch coverage over time, and the time spent in synchronization between the workers. Synchronization time has been measured based on the different policies used as follows. For each MPI-based configuration, the CPU cycles spent on sending and receiving messages are taken into account. More specifically, time is measured in \texttt{save\_if\_interesting}, where messages are sent to other fuzzers, and in \texttt{sync\_fuzzers}, where the \texttt{MPI\_Recv} loop has been placed.
For AFL configurations (both baseline and NFS-based), synchronization time is computed in \texttt{sync\_fuzzers}, counting CPU cycles employed to open files from the filesystem and save test cases into the fuzzers' owned directory.

\begin{figure}[!ht]
	\centering
	\begin{subfigure}[b]{\textwidth}
		\includegraphics[width=\textwidth]{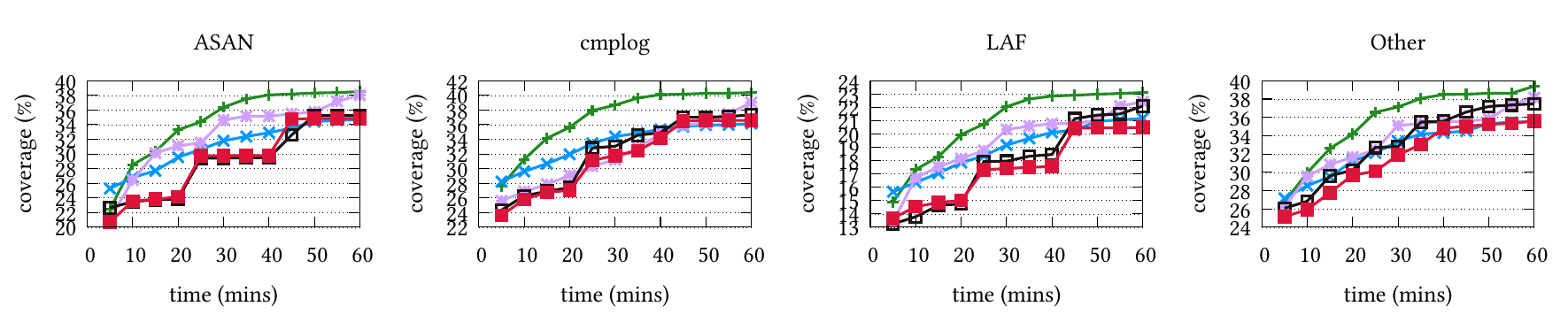}
		\includegraphics[width=\textwidth]{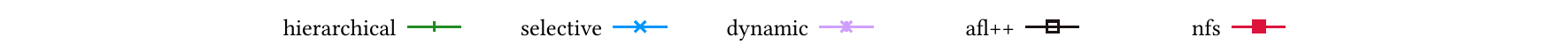}
		\acmcaption{Coverage.}
		\label{fig:freetype2-coverage}
	\end{subfigure}
	\begin{subfigure}[b]{\textwidth}
		\includegraphics[width=\textwidth]{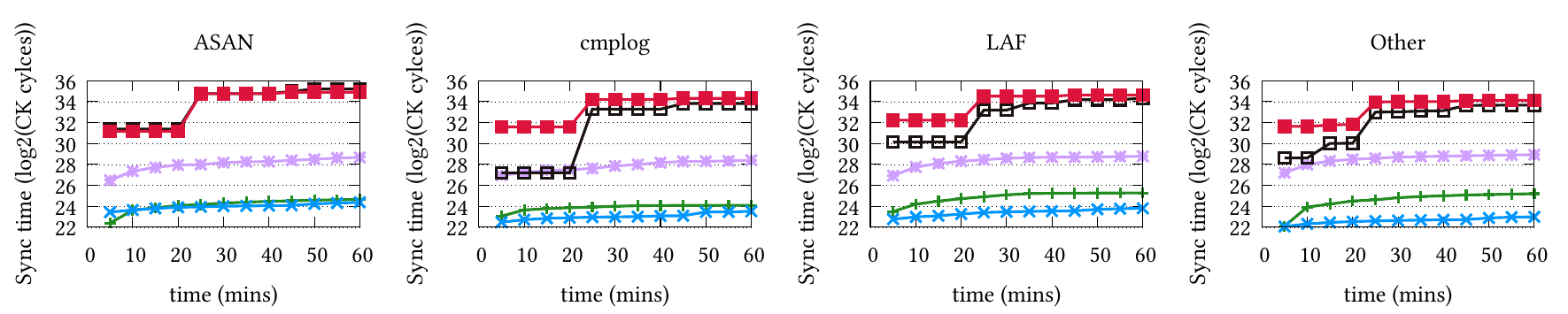}
		\includegraphics[width=\textwidth]{data/freetype2/key.pdf}
		\acmcaption{Synchronization Time.}
		\label{fig:freetype2-synctime}
	\end{subfigure}
	\begin{subfigure}[b]{\textwidth}
		\includegraphics[width=\textwidth]{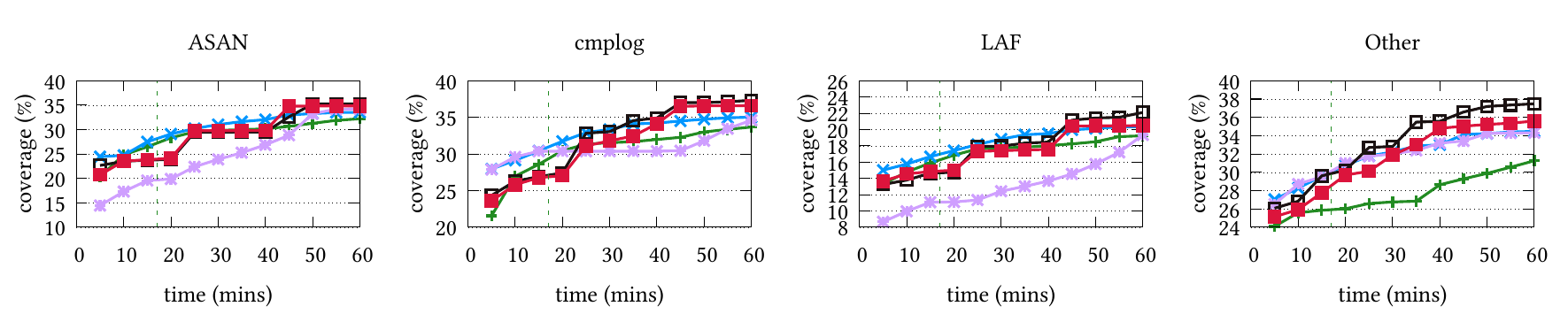}
		\includegraphics[width=\textwidth]{data/freetype2/key.pdf}
		\acmcaption{Coverage (with ammuina).}
		\label{fig:freetype2-coverage-ammuina}
	\end{subfigure}
	\begin{subfigure}[b]{\textwidth}
		\includegraphics[width=\textwidth]{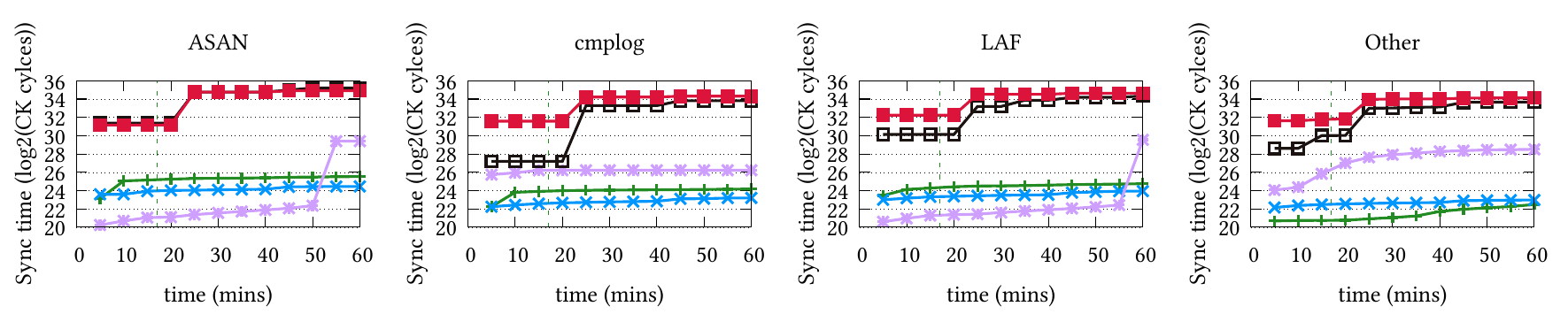}
		\includegraphics[width=\textwidth]{data/freetype2/key.pdf}
		\acmcaption{Synchronization Time (with ammuina).}
		\label{fig:freetype2-synctime-ammuina}
	\end{subfigure}
	\acmcaption{freetype2}
	\label{fig:freetype2}
\end{figure}

\begin{figure}[!ht]
	\centering
	\begin{subfigure}[b]{\textwidth}
		\includegraphics[width=\textwidth]{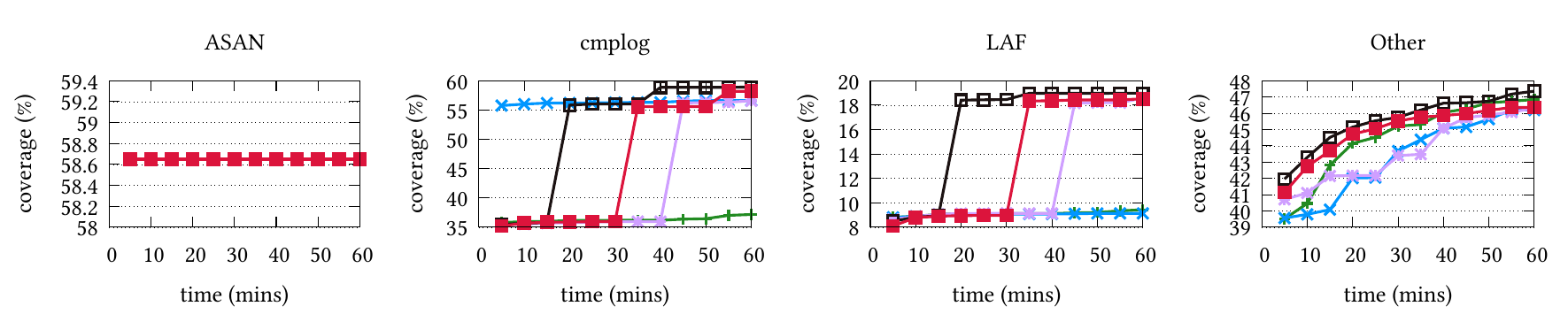}
		\includegraphics[width=\textwidth]{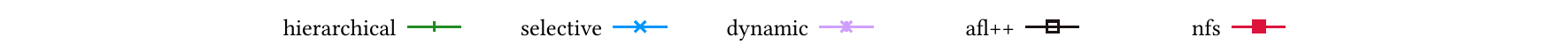}
		\acmcaption{Coverage.}
		\label{fig:guetzli-coverage}
	\end{subfigure}
	\begin{subfigure}[b]{\textwidth}
		\includegraphics[width=\textwidth]{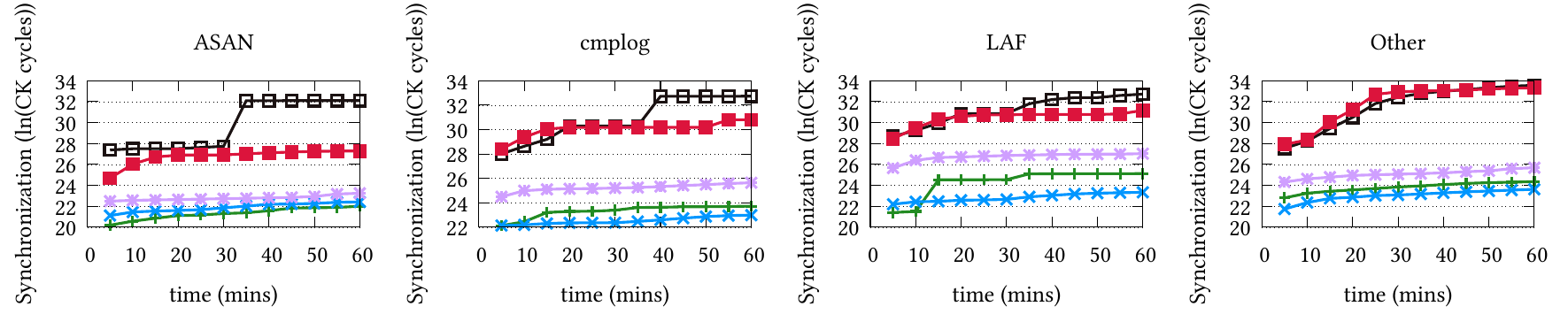}
		\includegraphics[width=\textwidth]{data/guetzli/key.pdf}
		\acmcaption{Synchronization Time.}
		\label{fig:guetzli-synctime}
	\end{subfigure}
	\begin{subfigure}[b]{\textwidth}
		\includegraphics[width=\textwidth]{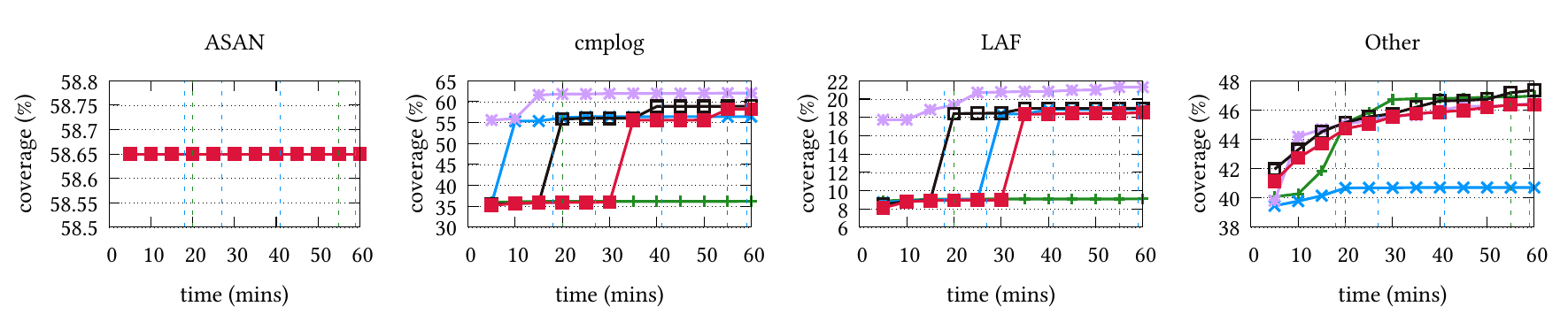}
		\includegraphics[width=\textwidth]{data/guetzli/key.pdf}
		\acmcaption{Coverage (with ammuina).}
		\label{fig:guetzli-coverage-ammuina}
	\end{subfigure}
	\begin{subfigure}[b]{\textwidth}
		\includegraphics[width=\textwidth]{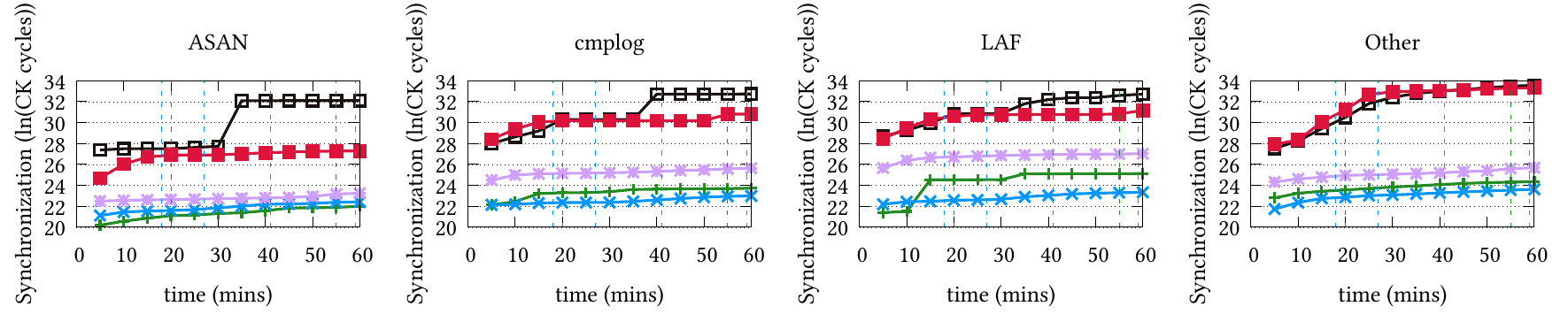}
		\includegraphics[width=\textwidth]{data/guetzli/key.pdf}
		\acmcaption{Synchronization Time (with ammuina).}
		\label{fig:guetzli-synctime-ammuina}
	\end{subfigure}
	\acmcaption{guetzli}
	\label{fig:guetzli}
\end{figure}

\begin{figure}[!ht]
	\centering
	\begin{subfigure}[b]{\textwidth}
		\includegraphics[width=\textwidth]{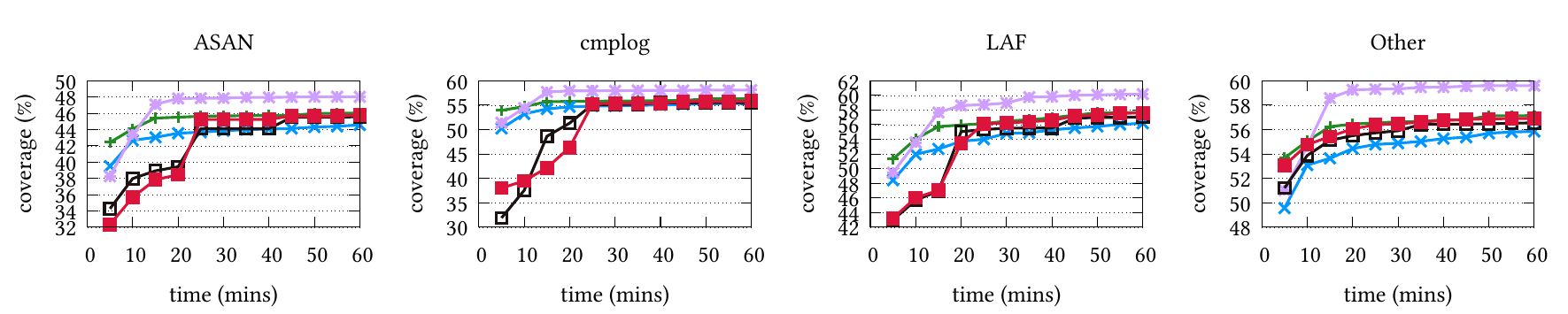}
		\includegraphics[width=\textwidth]{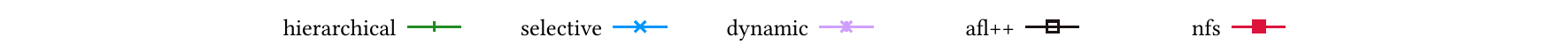}
		\acmcaption{Coverage.}
		\label{fig:harfbuzz-coverage}
	\end{subfigure}
	\begin{subfigure}[b]{\textwidth}
		\includegraphics[width=\textwidth]{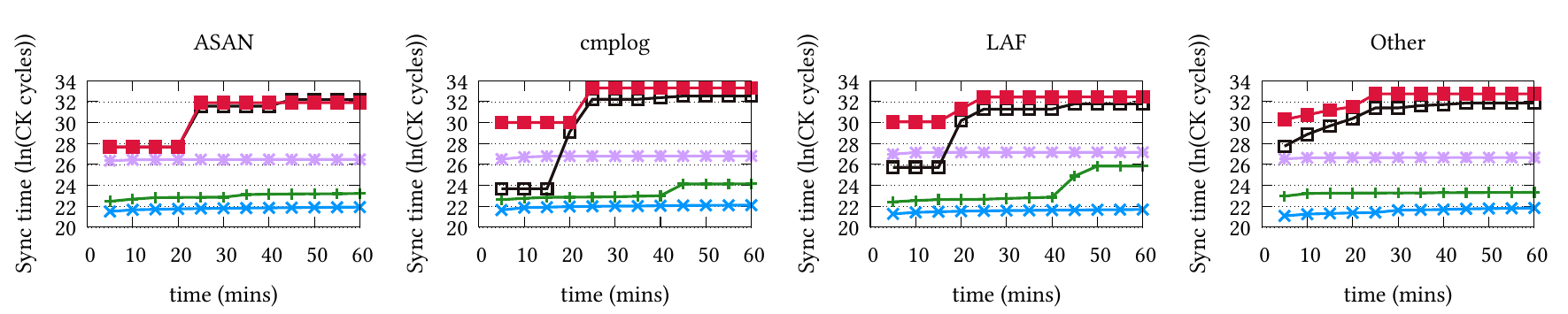}
		\includegraphics[width=\textwidth]{data/harfbuzz/key.pdf}
		\acmcaption{Synchronization Time.}
		\label{fig:harfbuzz-synctime}
	\end{subfigure}
	\begin{subfigure}[b]{\textwidth}
		\includegraphics[width=\textwidth]{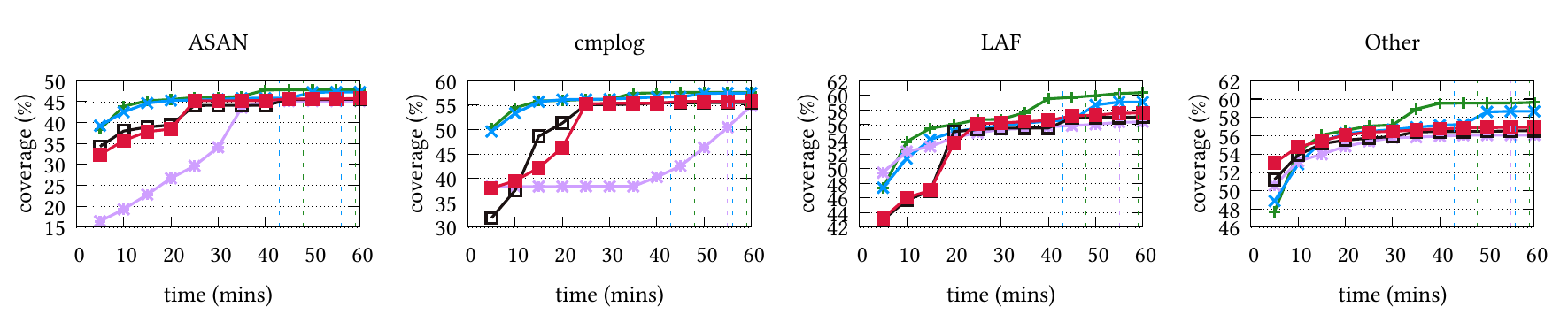}
		\includegraphics[width=\textwidth]{data/harfbuzz/key.pdf}
		\acmcaption{Coverage (with ammuina).}
		\label{fig:harfbuzz-coverage-ammuina}
	\end{subfigure}
	\begin{subfigure}[b]{\textwidth}
		\includegraphics[width=\textwidth]{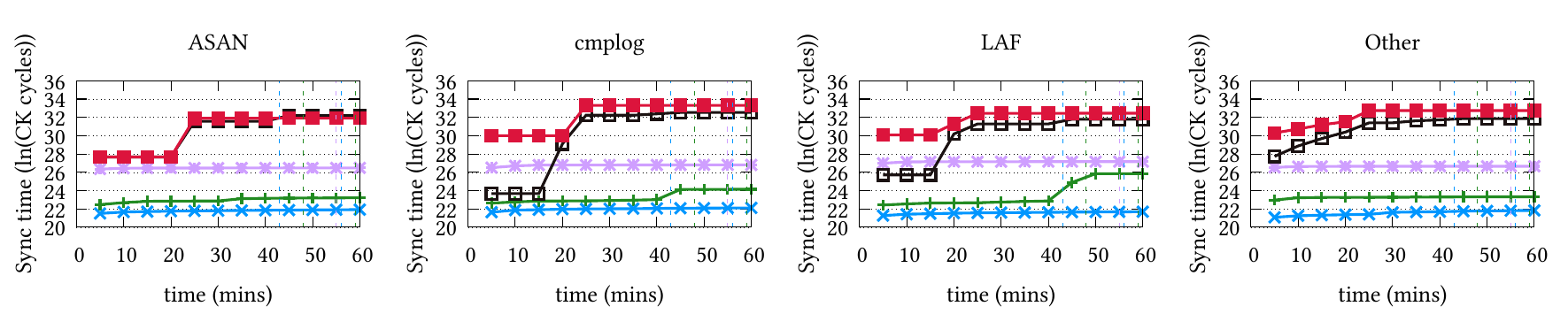}
		\includegraphics[width=\textwidth]{data/harfbuzz/key.pdf}
		\acmcaption{Synchronization Time (with ammuina).}
		\label{fig:harfbuzz-synctime-ammuina}
	\end{subfigure}
	\acmcaption{harfbuzz}
	\label{fig:harfbuzz}
\end{figure}

\begin{figure}[!ht]
	\centering
	\begin{subfigure}[b]{\textwidth}
		\includegraphics[width=\textwidth]{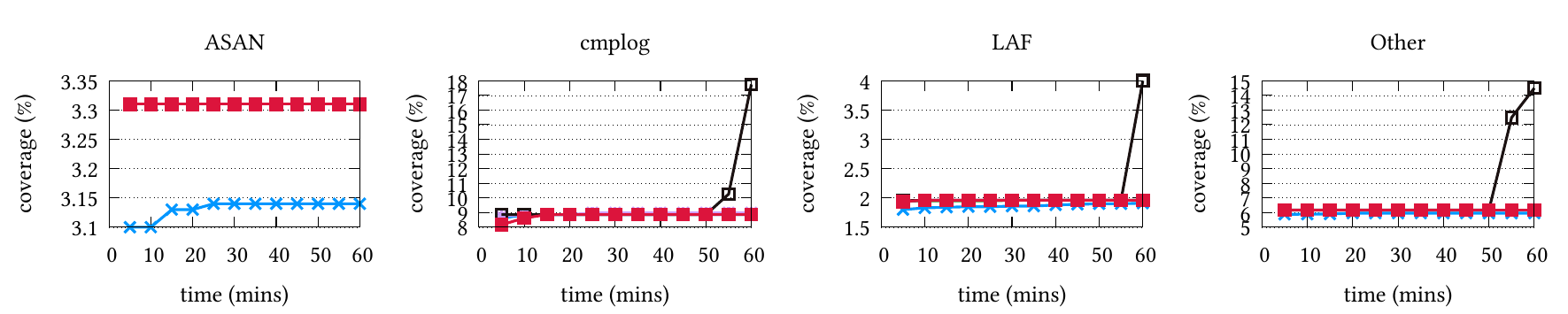}
		\includegraphics[width=\textwidth]{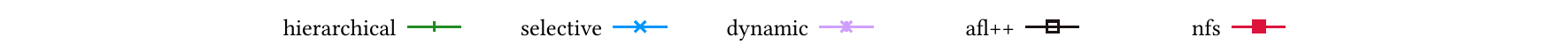}
		\acmcaption{Coverage.}
		\label{fig:lcms-coverage}
	\end{subfigure}
	\begin{subfigure}[b]{\textwidth}
		\includegraphics[width=\textwidth]{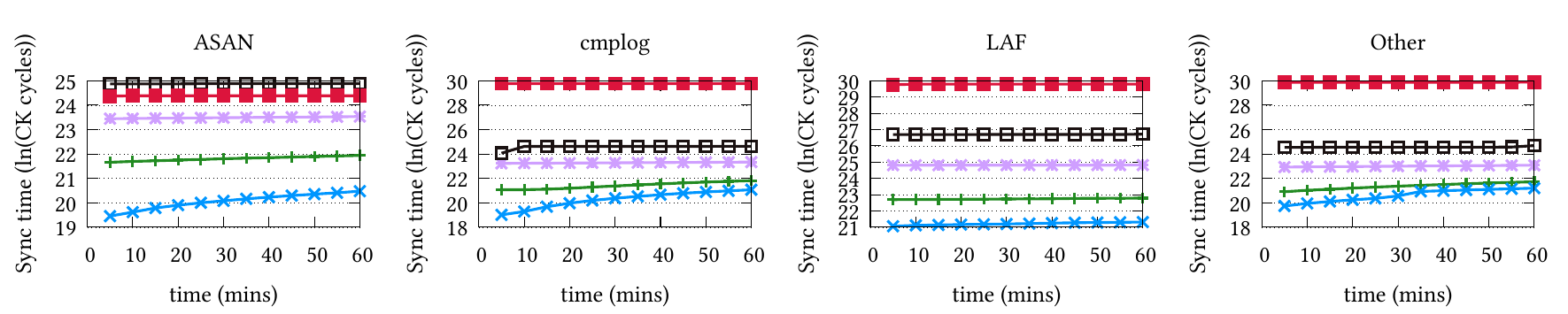}
		\includegraphics[width=\textwidth]{data/lcms/key.pdf}
		\acmcaption{Synchronization Time.}
		\label{fig:lcms-synctime}
	\end{subfigure}
	\begin{subfigure}[b]{\textwidth}
		\includegraphics[width=\textwidth]{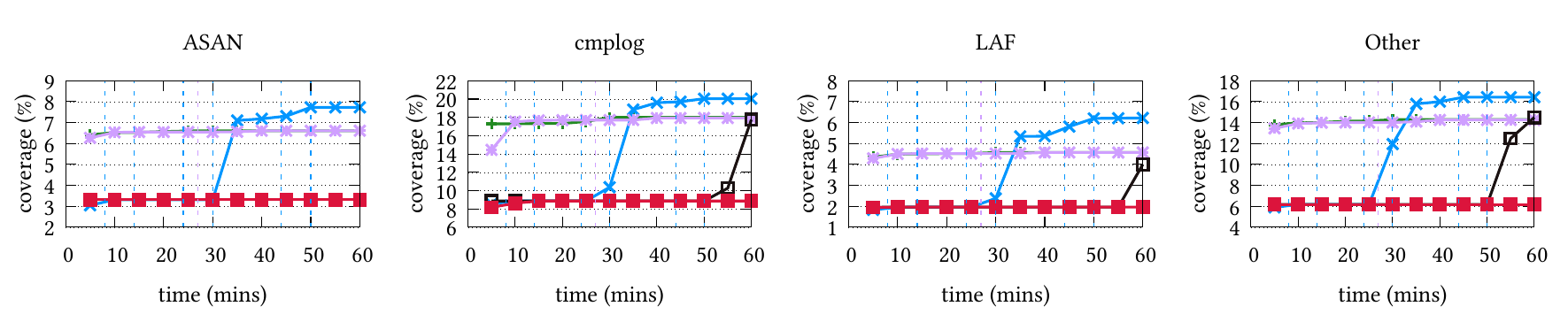}
		\includegraphics[width=\textwidth]{data/lcms/key.pdf}
		\acmcaption{Coverage (with ammuina).}
		\label{fig:lcms-coverage-ammuina}
	\end{subfigure}
	\begin{subfigure}[b]{\textwidth}
		\includegraphics[width=\textwidth]{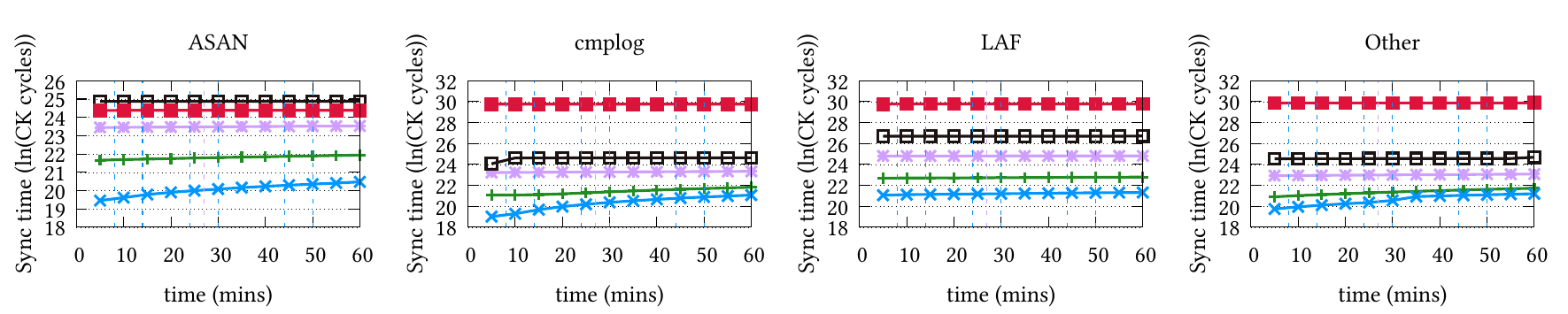}
		\includegraphics[width=\textwidth]{data/lcms/key.pdf}
		\acmcaption{Synchronization Time (with ammuina).}
		\label{fig:lcms-synctime-ammuina}
	\end{subfigure}
	\acmcaption{lcms}
	\label{fig:lcms}
\end{figure}

\begin{figure}[!ht]
    \centering
    \begin{subfigure}[b]{\textwidth}
        \includegraphics[width=\textwidth]{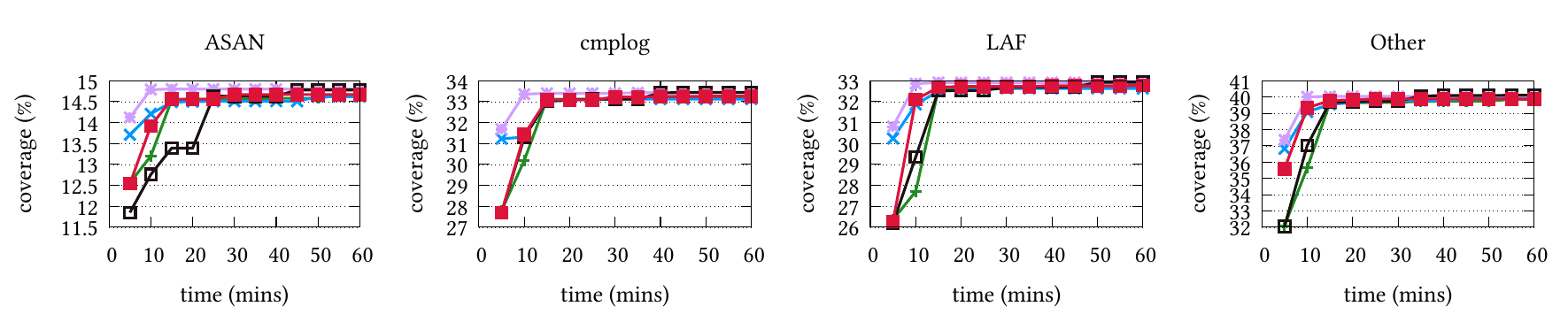}
        \includegraphics[width=\textwidth]{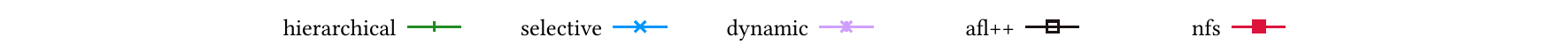}
        \acmcaption{Coverage.}
        \label{fig:libjpeg-coverage}
    \end{subfigure}
    \begin{subfigure}[b]{\textwidth}
        \includegraphics[width=\textwidth]{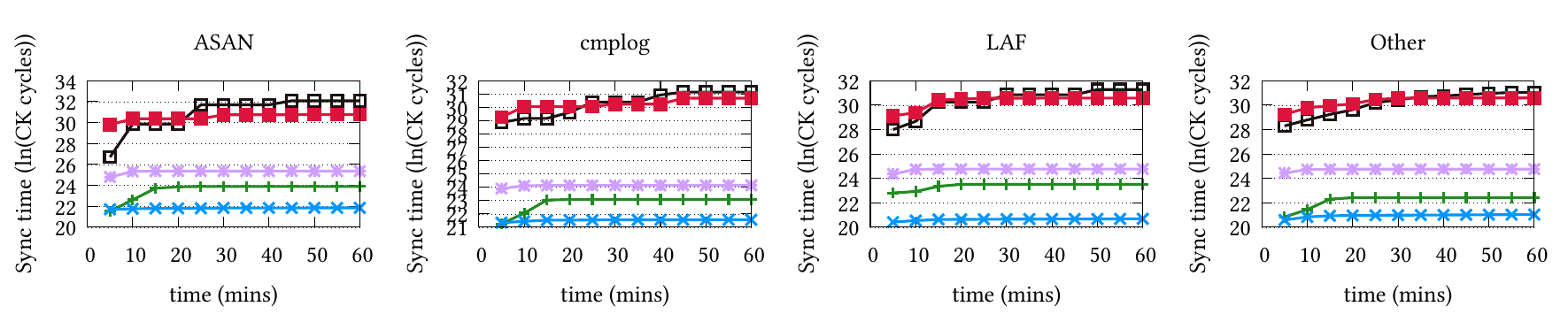}
        \includegraphics[width=\textwidth]{data/libjpeg/key.pdf}
        \acmcaption{Synchronization Time.}
        \label{fig:libjpeg-synctime}
    \end{subfigure}
    \begin{subfigure}[b]{\textwidth}
    	\includegraphics[width=\textwidth]{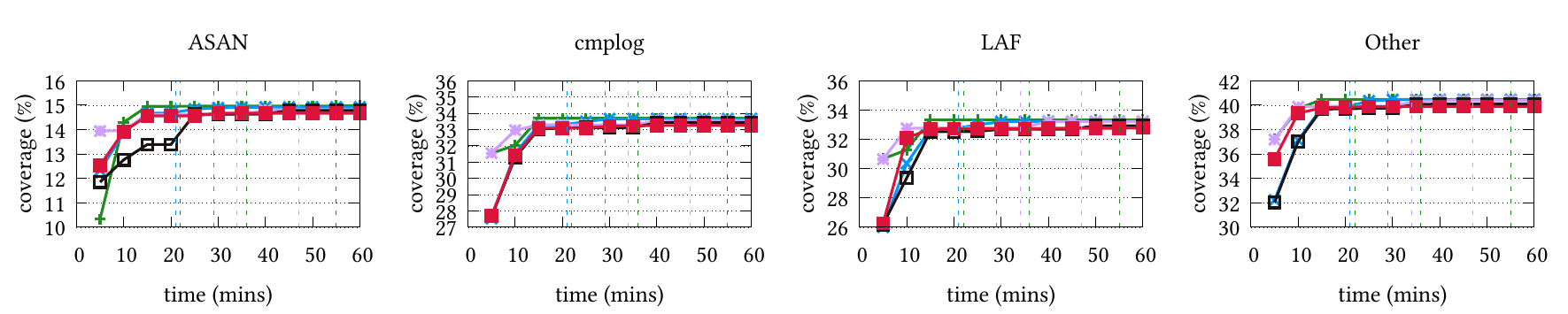}
    	\includegraphics[width=\textwidth]{data/libjpeg/key.pdf}
    	\acmcaption{Coverage (with ammuina).}
    	\label{fig:libjpeg-coverage-ammuina}
    \end{subfigure}
    \begin{subfigure}[b]{\textwidth}
    	\includegraphics[width=\textwidth]{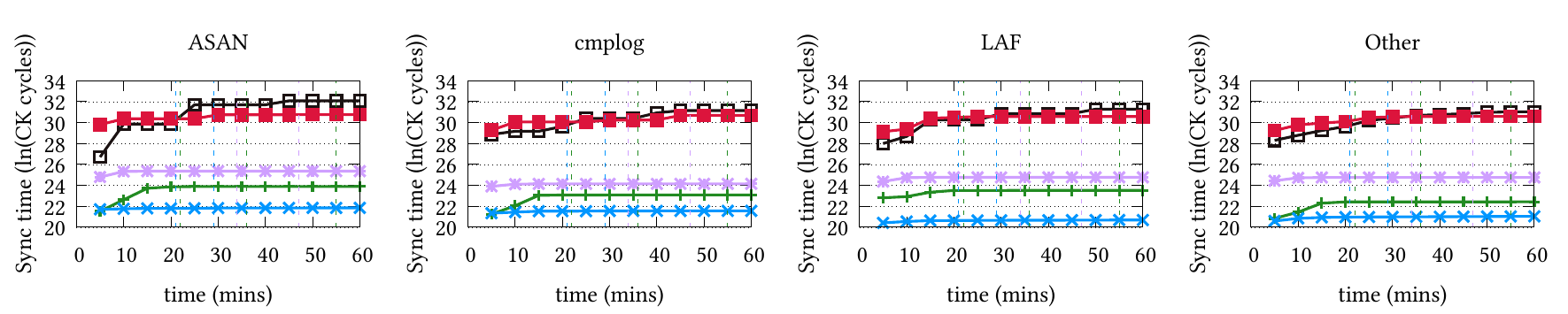}
    	\includegraphics[width=\textwidth]{data/libjpeg/key.pdf}
    	\acmcaption{Synchronization Time (with ammuina).}
    	\label{fig:libjpeg-synctime-ammuina}
    \end{subfigure}
    \acmcaption{libjpeg}
    \label{fig:libjpeg}
\end{figure}

\begin{figure}[!ht]
	\centering
	\begin{subfigure}[b]{\textwidth}
		\includegraphics[width=\textwidth]{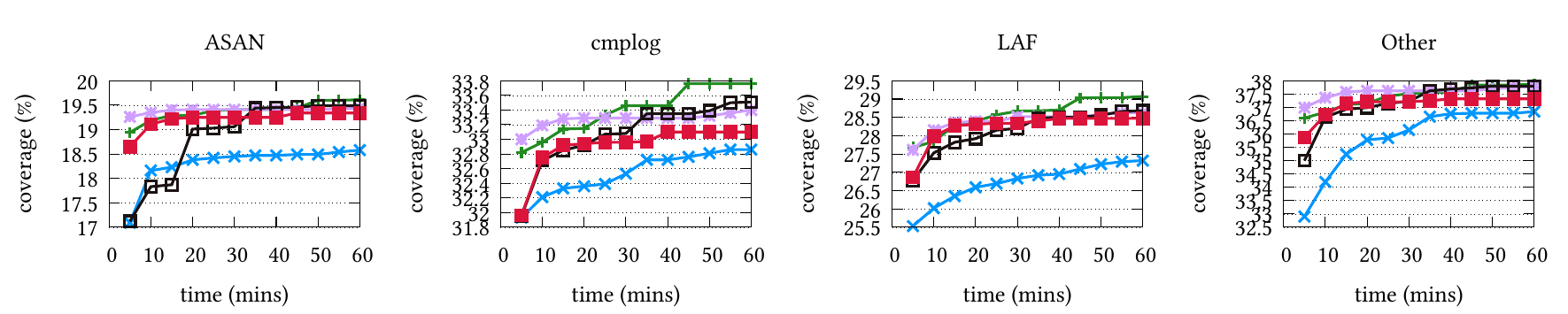}
		\includegraphics[width=\textwidth]{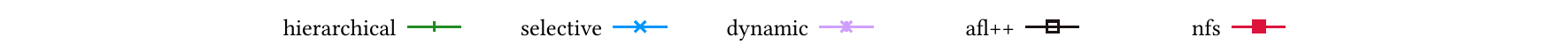}
		\acmcaption{Coverage.}
		\label{fig:libpng-coverage}
	\end{subfigure}
	\begin{subfigure}[b]{\textwidth}
		\includegraphics[width=\textwidth]{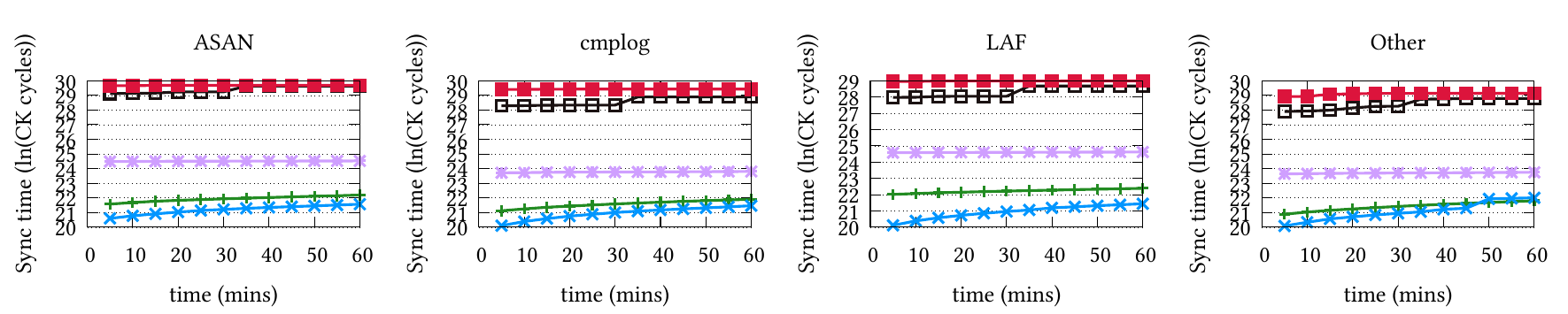}
		\includegraphics[width=\textwidth]{data/libpng/key.pdf}
		\acmcaption{Synchronization Time.}
		\label{fig:libpng-synctime}
	\end{subfigure}
	\begin{subfigure}[b]{\textwidth}
		\includegraphics[width=\textwidth]{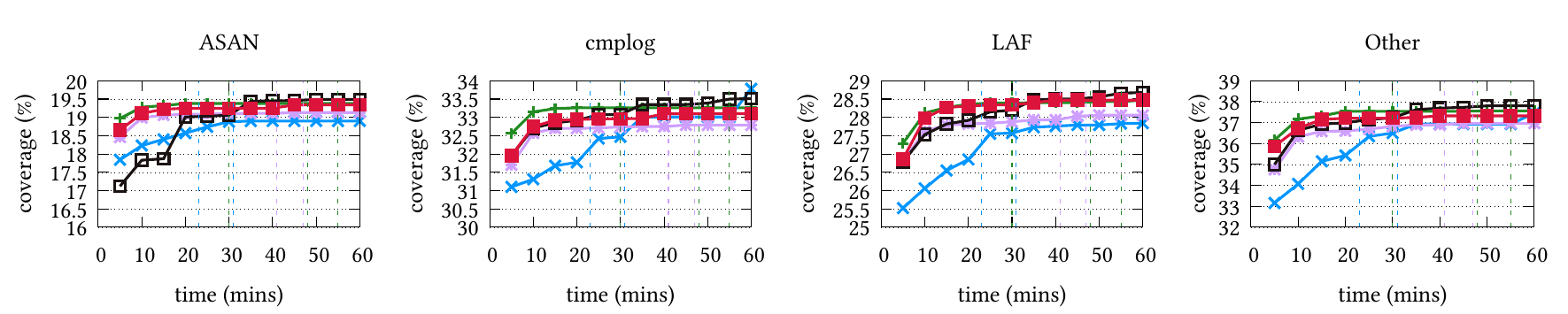}
		\includegraphics[width=\textwidth]{data/libpng/key.pdf}
		\acmcaption{Coverage (with ammuina).}
		\label{fig:libpng-coverage-ammuina}
	\end{subfigure}
	\begin{subfigure}[b]{\textwidth}
		\includegraphics[width=\textwidth]{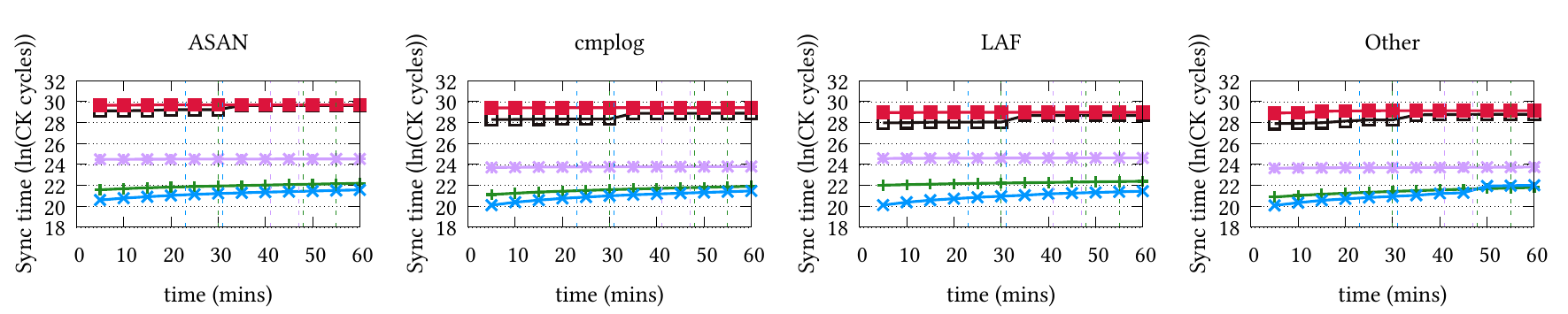}
		\includegraphics[width=\textwidth]{data/libpng/key.pdf}
		\acmcaption{Synchronization Time (with ammuina).}
		\label{fig:libpng-synctime-ammuina}
	\end{subfigure}
	\acmcaption{libpng}
	\label{fig:libpng}
\end{figure}

\begin{figure}[!ht]
	\centering
	\begin{subfigure}[b]{\textwidth}
		\includegraphics[width=\textwidth]{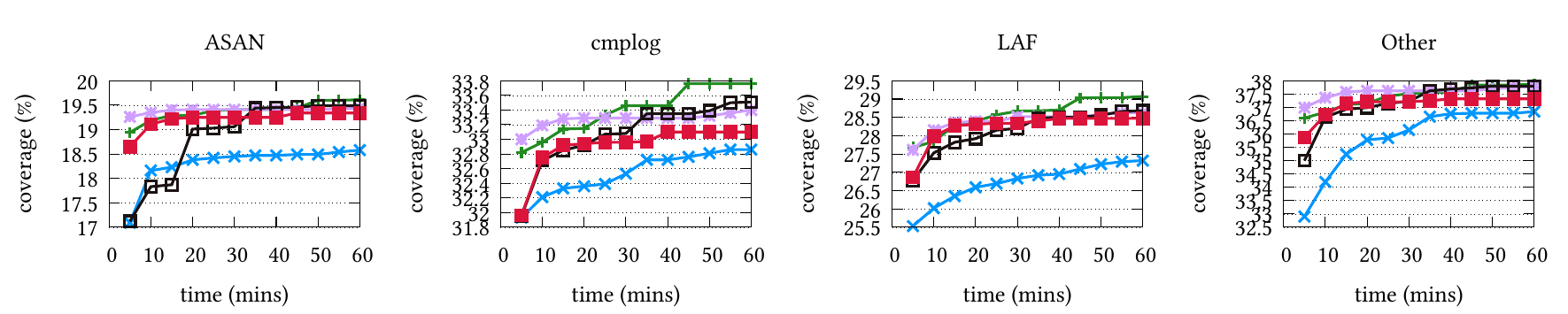}
		\includegraphics[width=\textwidth]{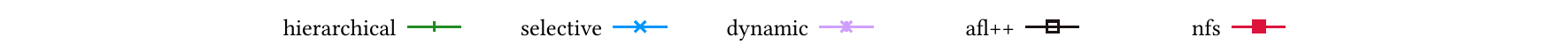}
		\acmcaption{Coverage.}
		\label{fig:pcre2-coverage}
	\end{subfigure}
	\begin{subfigure}[b]{\textwidth}
		\includegraphics[width=\textwidth]{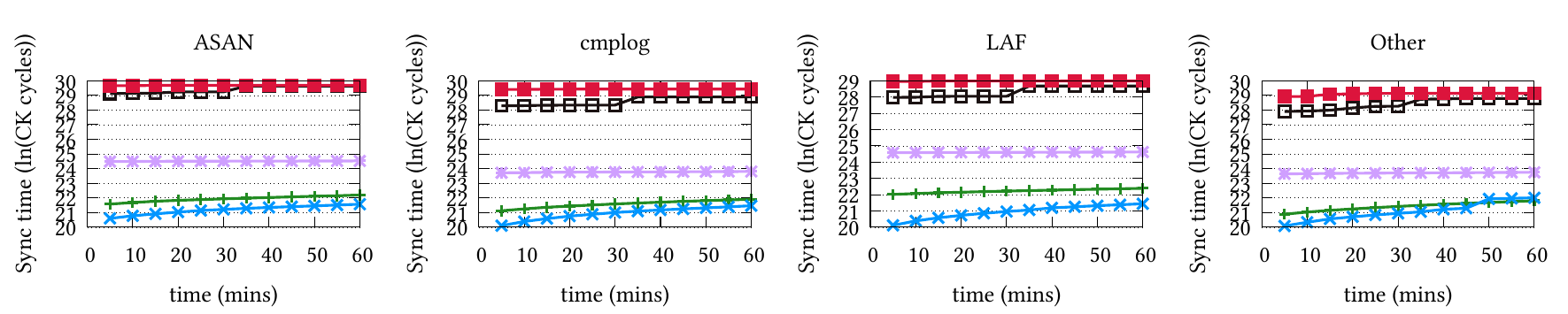}
		\includegraphics[width=\textwidth]{data/pcre2/key.pdf}
		\acmcaption{Synchronization Time.}
		\label{fig:pcre2-synctime}
	\end{subfigure}
	\begin{subfigure}[b]{\textwidth}
		\includegraphics[width=\textwidth]{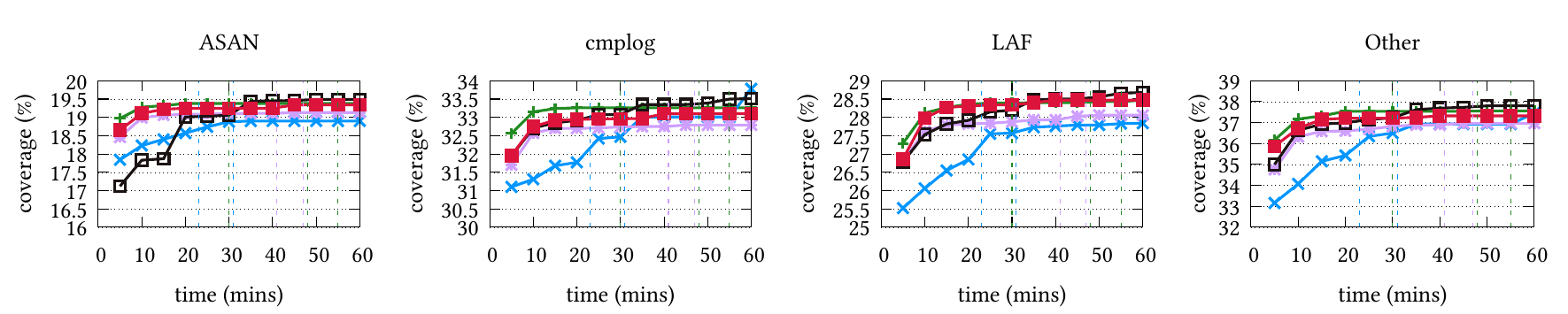}
		\includegraphics[width=\textwidth]{data/pcre2/key.pdf}
		\acmcaption{Coverage (with ammuina).}
		\label{fig:pcre2-coverage-ammuina}
	\end{subfigure}
	\begin{subfigure}[b]{\textwidth}
		\includegraphics[width=\textwidth]{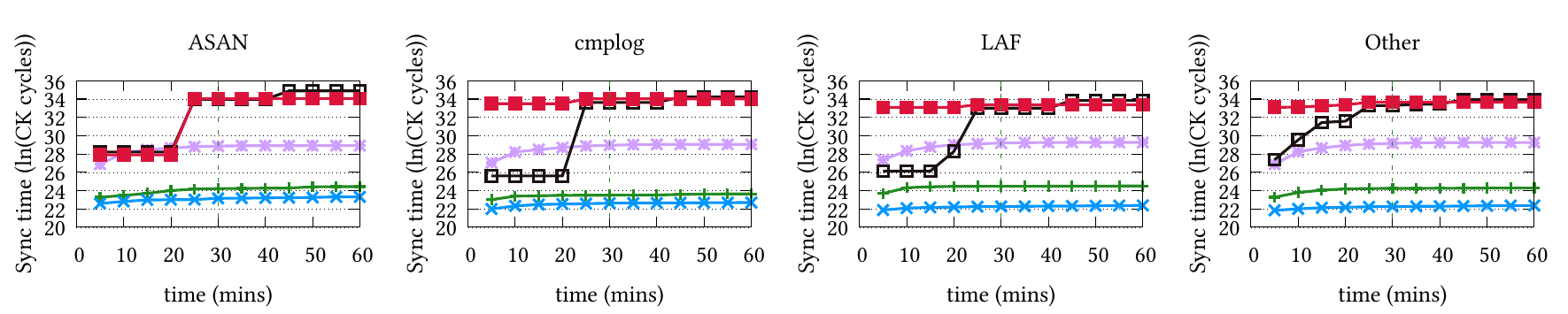}
		\includegraphics[width=\textwidth]{data/pcre2/key.pdf}
		\acmcaption{Synchronization Time (with ammuina).}
		\label{fig:pcre2-synctime-ammuina}
	\end{subfigure}
	\acmcaption{pcre2}
	\label{fig:pcre2}
\end{figure}

\begin{figure}[!ht]
	\centering
	\begin{subfigure}[b]{\textwidth}
		\includegraphics[width=\textwidth]{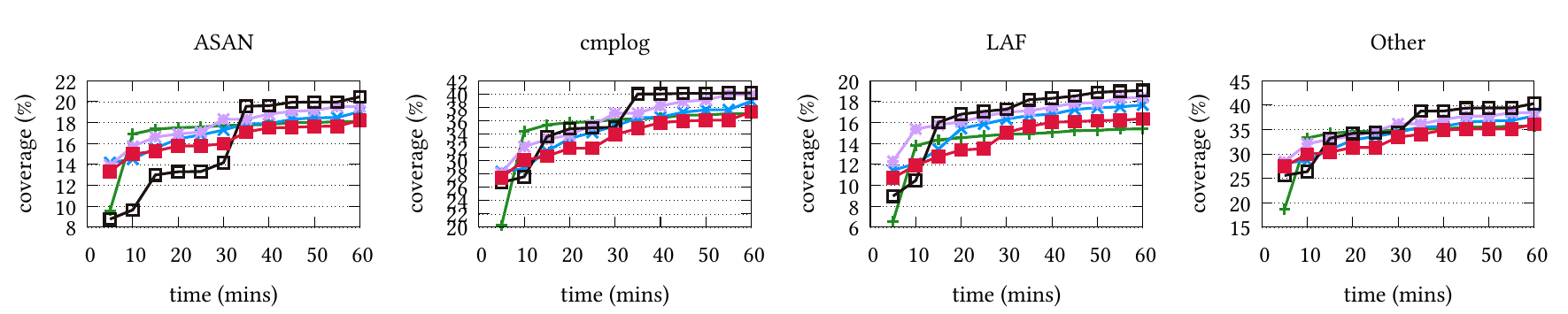}
		\includegraphics[width=\textwidth]{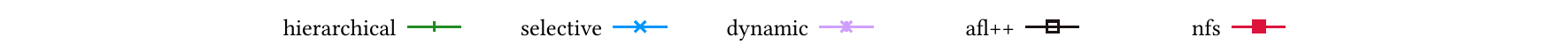}
		\acmcaption{Coverage.}
		\label{fig:proj4-coverage}
	\end{subfigure}
	\begin{subfigure}[b]{\textwidth}
		\includegraphics[width=\textwidth]{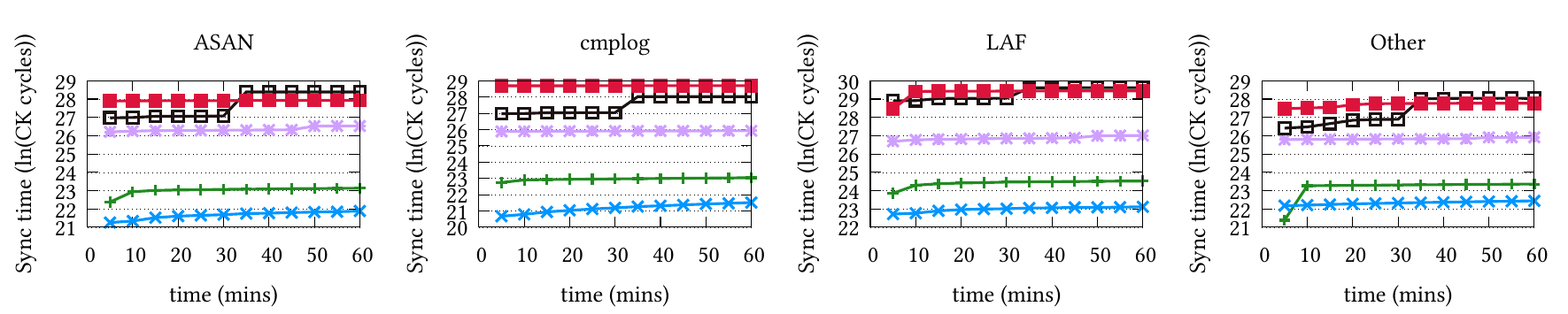}
		\includegraphics[width=\textwidth]{data/proj4/key.pdf}
		\acmcaption{Synchronization Time.}
		\label{fig:proj4-synctime}
	\end{subfigure}
	\begin{subfigure}[b]{\textwidth}
		\includegraphics[width=\textwidth]{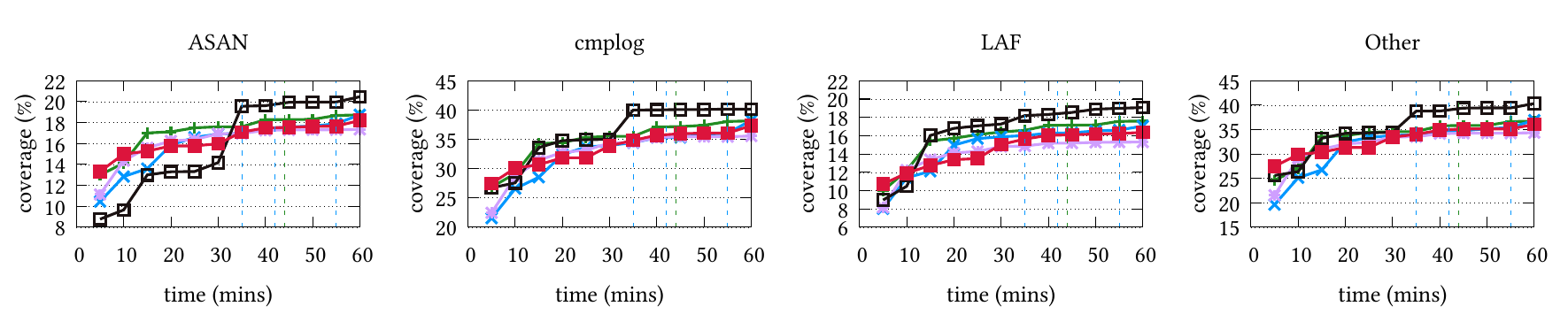}
		\includegraphics[width=\textwidth]{data/proj4/key.pdf}
		\acmcaption{Coverage (with ammuina).}
		\label{fig:proj4-coverage-ammuina}
	\end{subfigure}
	\begin{subfigure}[b]{\textwidth}
		\includegraphics[width=\textwidth]{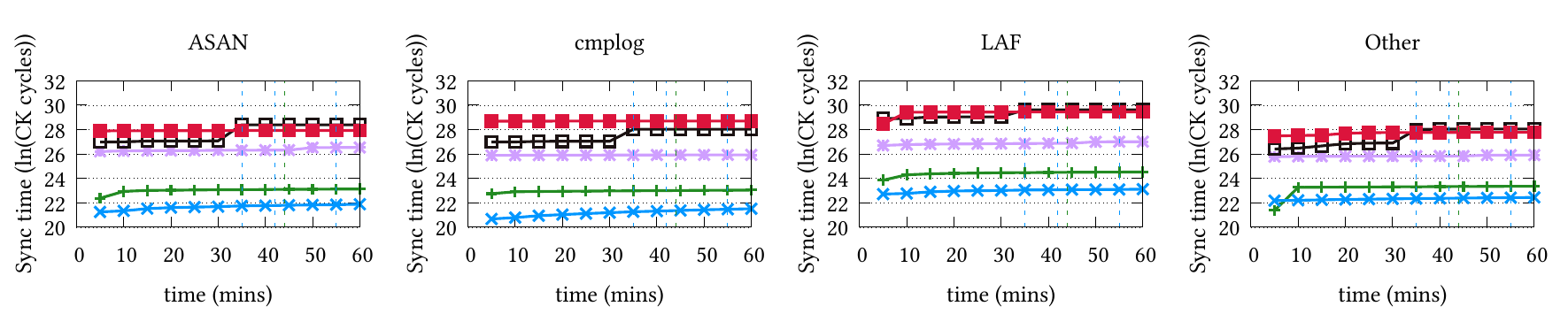}
		\includegraphics[width=\textwidth]{data/proj4/key.pdf}
		\acmcaption{Synchronization Time (with ammuina).}
		\label{fig:proj4-synctime-ammuina}
	\end{subfigure}
	\acmcaption{proj4}
	\label{fig:proj4}
\end{figure}

\begin{figure}[!ht]
	\centering
	\begin{subfigure}[b]{\textwidth}
		\includegraphics[width=\textwidth]{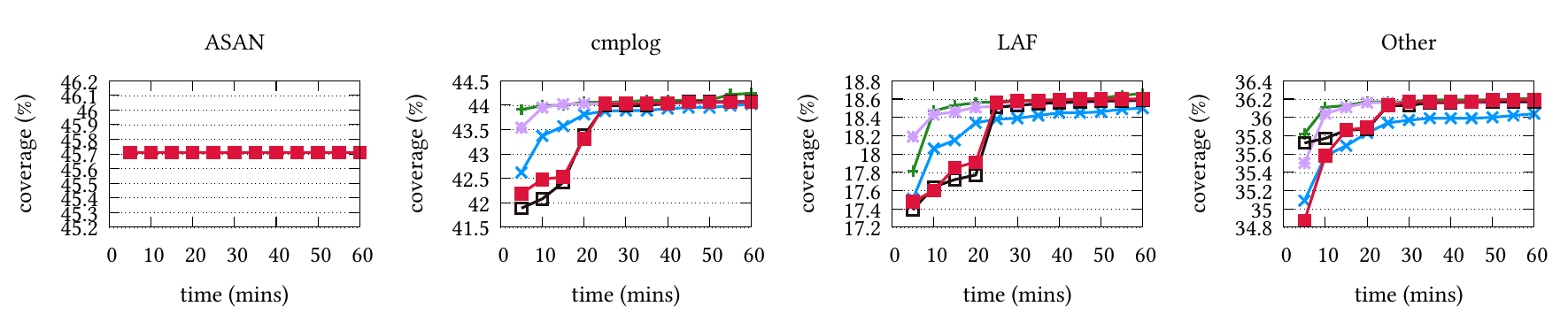}
		\includegraphics[width=\textwidth]{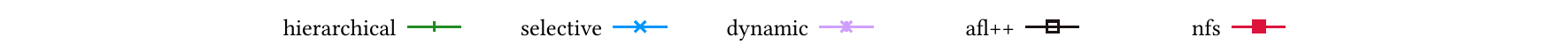}
		\acmcaption{Coverage.}
		\label{fig:re2-coverage}
	\end{subfigure}
	\begin{subfigure}[b]{\textwidth}
		\includegraphics[width=\textwidth]{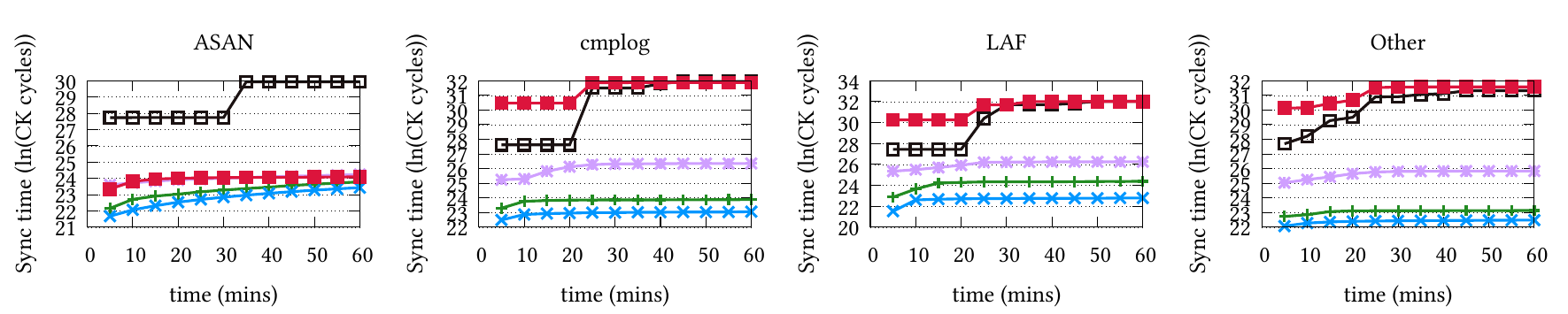}
		\includegraphics[width=\textwidth]{data/re2/key.pdf}
		\acmcaption{Synchronization Time.}
		\label{fig:re2-synctime}
	\end{subfigure}
	\begin{subfigure}[b]{\textwidth}
		\includegraphics[width=\textwidth]{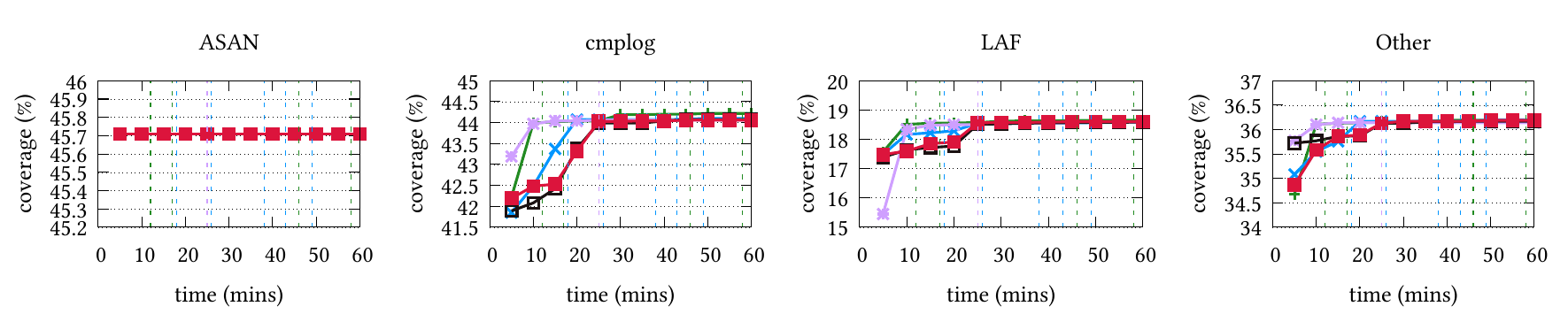}
		\includegraphics[width=\textwidth]{data/re2/key.pdf}
		\acmcaption{Coverage (with ammuina).}
		\label{fig:re2-coverage-ammuina}
	\end{subfigure}
	\begin{subfigure}[b]{\textwidth}
		\includegraphics[width=\textwidth]{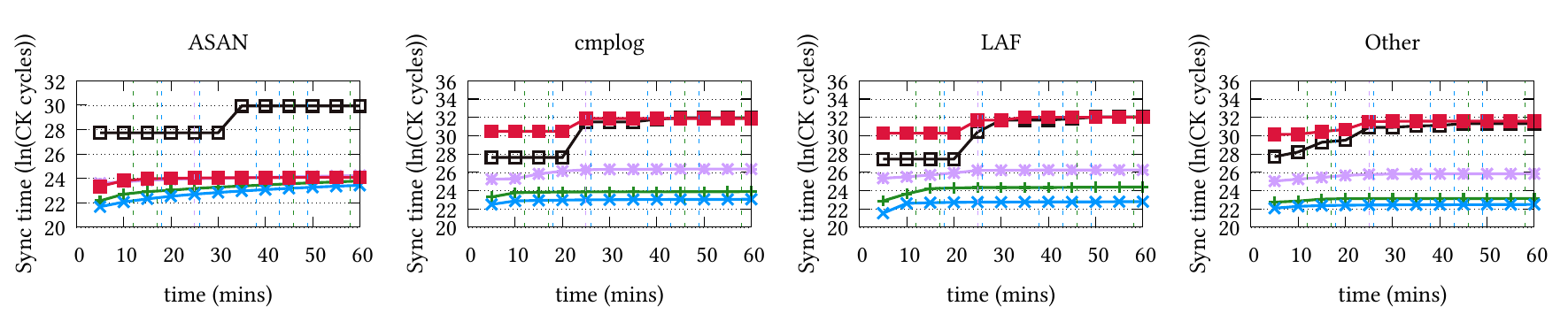}
		\includegraphics[width=\textwidth]{data/re2/key.pdf}
		\acmcaption{Synchronization Time (with ammuina).}
		\label{fig:re2-synctime-ammuina}
	\end{subfigure}
	\acmcaption{re2}
	\label{fig:re2}
\end{figure}

\begin{figure}[!ht]
	\centering
	\begin{subfigure}[b]{\textwidth}
		\includegraphics[width=\textwidth]{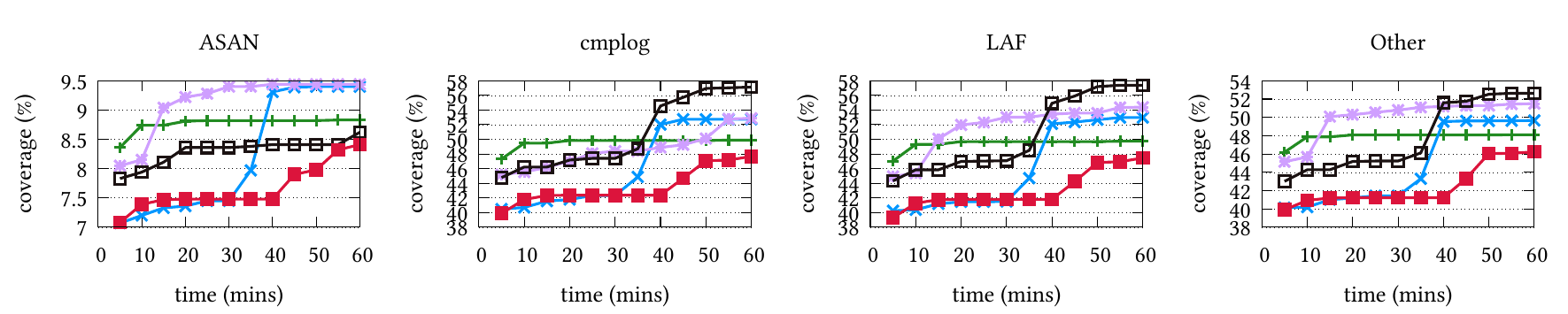}
		\includegraphics[width=\textwidth]{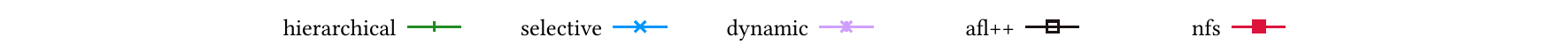}
		\acmcaption{Coverage.}
		\label{fig:woff2-coverage}
	\end{subfigure}
	\begin{subfigure}[b]{\textwidth}
		\includegraphics[width=\textwidth]{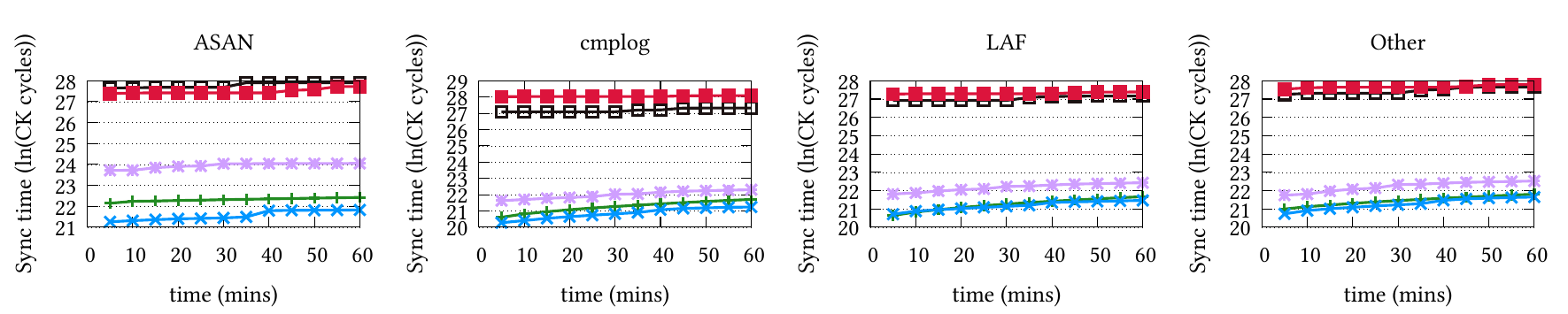}
		\includegraphics[width=\textwidth]{data/woff2/key.pdf}
		\acmcaption{Synchronization Time.}
		\label{fig:woff2-synctime}
	\end{subfigure}
	\begin{subfigure}[b]{\textwidth}
		\includegraphics[width=\textwidth]{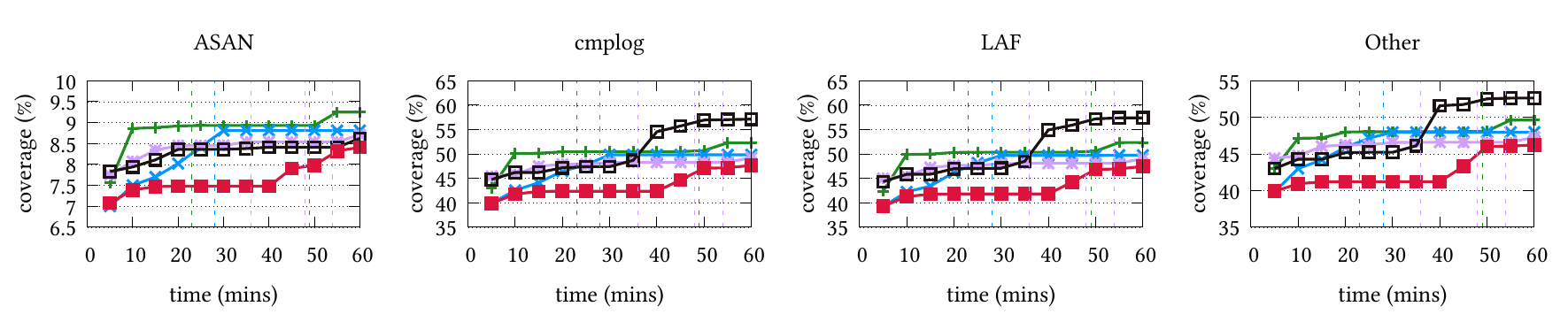}
		\includegraphics[width=\textwidth]{data/woff2/key.pdf}
		\acmcaption{Coverage (with ammuina).}
		\label{fig:woff2-coverage-ammuina}
	\end{subfigure}
	\begin{subfigure}[b]{\textwidth}
		\includegraphics[width=\textwidth]{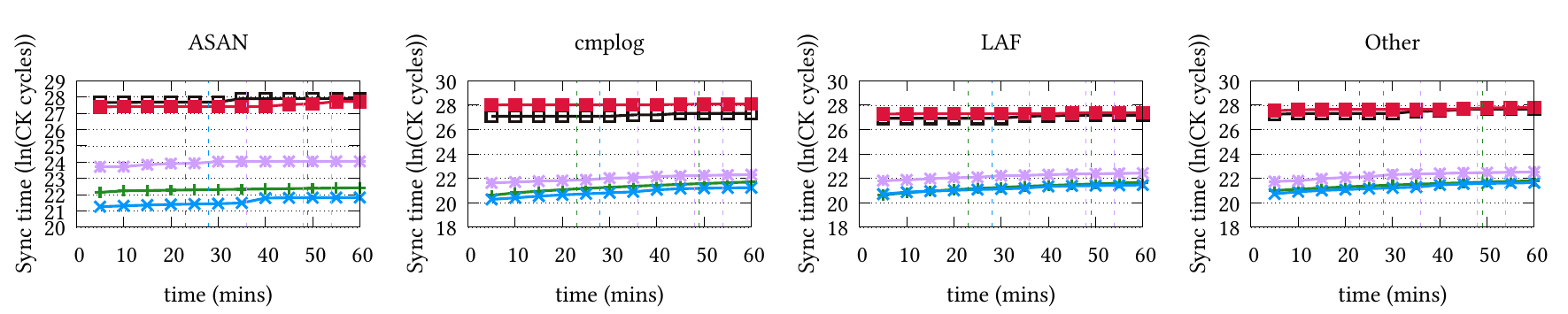}
		\includegraphics[width=\textwidth]{data/woff2/key.pdf}
		\acmcaption{Synchronization Time (with ammuina).}
		\label{fig:woff2-synctime-ammuina}
	\end{subfigure}
	\acmcaption{woff2}
	\label{fig:woff2}
\end{figure}

Figures~\ref{fig:freetype2}--\ref{fig:woff2} provide the results for all the considered benchmarks. As can be seen from the results, most of the benchmarks show a more timely increase in the coverage when employing some MPI-based dissemination policy, particularly for the cases of freetype2 (Figure~\ref{fig:freetype2-coverage}), harfbuzz (Figure~\ref{fig:harfbuzz-coverage}), libjpeg (Figure~\ref{fig:libjpeg-coverage}), libpng (Figure~\ref{fig:libpng-coverage}), pcre2 (Figure~\ref{fig:pcre2-coverage}), and woff2 (Figure~\ref{fig:woff2-coverage}), although the specific policy that delivers the timeliest coverage differs.

For the largest part of the considered benchmarks, the hierarchical policy delivers a higher coverage very quickly, i.e. after 10-15 minutes of fuzzing it is able to outperform other strategies---notably, in the case of freetype2, as reported in Figure~\ref{fig:freetype2-coverage}, the hierarchical policy also consistently outperforms any other policy with any fuzzing configuration, although this result holds only for this very specific benchmark application. At the same time, we observe that the dynamic policy can quickly regain pace and become the fastest policy at delivering a higher coverage in a limited time. For example, in the case of the woff2 benchmark (see Figure~\ref{fig:woff2-coverage}), ASAN, LAF, and other fuzzers show a trend in coverage such that after 15--20 minutes, the dynamic policy provides better coverage results.

This behaviour of the dynamic policy is expected: since this policy is based on utilization, it attempts to favor the dissemination of new inputs to the fuzzing nodes that are able to make the best use of it. Nevertheless, there are cases in which this policy takes more time to show a notable improvement. For example, in the case of woff2 using cmplog (see again Figure~\ref{fig:woff2-coverage}), proj4 using cmplog (see Figure~\ref{fig:proj4-coverage}), or pcre2 (see Figure~\ref{fig:pcre2-coverage}, the dynamic policy requires more time to start outperforming the others. In these cases, though, the hierarchical policy shows a fast increase in coverage. The reason behind this result can be related to the organization of the fuzzing network, where a tighter interconnection between groups of the same fuzzers can increase the likelihood that useful inputs are received more promptly.

The reason behind the improvement of MPI-based policies is also fundamentally tied to the drastically reduced synchronization latency achieved by replacing filesystem‐based checkpoints with lightweight MPI point‐to‐point operations. Rather than opening, reading, and writing files on disk or over NFS, each rank dispatches ``interesting'' test cases via non-blocking \texttt{MPI\_Isend} calls and retrieves them with \texttt{MPI\_Recv} in a loop; the total CPU cycles devoted to synchronization under MPI‐based policies are rarely dominant, while AFL++’s scp-based synchronization and NFS mounts incur larger overheads. This reduction is especially evident when fuzzing ASAN-instrumented binaries, where the heavier code-coverage feedback amplifies filesystem stalls.

In general, the performance of lcms (see Figure~\ref{fig:lcms}) in the distributed fuzzing setup is demanding: code coverage increases slowly, and all synchronization policies struggle to make meaningful advances. The baseline policies also face similar challenges, although they achieve slightly better performance. For instance, in the cases of cmplog, LAF, and Others, the baseline AFL++ shows a sudden increase in coverage.

At the same time, this benchmark illustrates well how applying ammuina can yield tangible benefits. When ammuina is enabled, the MPI-based policies experience a noticeable improvement in the rate of coverage growth. The most compelling evidence comes from observing the selective policy under ASAN instrumentation. Although this configuration triggers a large number of ammuina events (much more frequently than other setups, also leading to a slight increase in synchronization time over time), it still initially struggles to gain traction.

However, the cumulative small benefits from each ammuina event gradually build enough momentum to overcome the initial plateau. As a result, the policy eventually catches up, exits stagnation, and enters a phase of progress. This delayed yet significant turnaround highlights that while ammuina can be effective even under unfavorable conditions, its impact may only become apparent after achieving a critical mass of input diversity across the various fuzzing ranks.

In any case, ammuina is unable to speed up program exploration under all circumstances. For example, in the case of proj4 (see Figure~\ref{fig:proj4-coverage-ammuina}), even though the hierarchical and selective policies trigger ammuina after some 30--40 minutes of execution, the coverage does not improve significantly compared to an execution without ammuina (Figure~\ref{fig:proj4-coverage}. In this particular case, the reason is related to the quality of exchanged inputs: they are not useful for the other fuzzers, thus bringing no significant advantage.

In the case of re2 (Figure~\ref{fig:re2-coverage}), we can observe that, for applications that are easy to fuzz---note that with ASAN all fuzzing configurations reach a steady state almost mmediately---the utilization of the MPI-based policies that we have proposed allow to reach a steady state (in the case of cmplog, LAF and Other) much more quickly. This is exactly the reason for the introduction of differentiated input exchange policies: groups of fuzzers that may be slower at exploring an application can benefit from different input exchanges. A similar behaviour can also be observed in the case of woff2 (see Figure~\ref{fig:woff2-coverage}): our MPI-based policies allow a faster exploration at the beginning of the fuzzing campaign, although after some time, more traditional fuzzing strategies (namely, plain AFL++) outperform them. In this particular case, ammuina is unable to improve the exploration for the reasons discussed above.

\begin{table}[t]
    \centering
    \begin{subtable}[t]{\linewidth}
    \resizebox{\linewidth}{!}{%
    \begin{tabular}{lcccccccccc}
    \toprule
    \textbf{Policy} & \textbf{freetype2} & \textbf{guetzli} & \textbf{harfbuzz} & \textbf{lcms} & \textbf{libjpeg} & \textbf{libpng} & \textbf{pcre2} & \textbf{proj4} & \textbf{re2} & \textbf{woff2}\\
    \midrule

    selective & \best{10} & \best{5} & \best{5} & --- & \best{5} & 20 & \best{5} & 15 & \best{5} & 40\\
    
    hierarchical & \best{10} & \best{5} & \best{5} & --- & 10 & \best{5} & \best{5} & \best{10} & \best{5} & \best{5}\\
    
    dynamic & 15 & \best{5} & \best{5} & --- & \best{5} & \best{5} & 10 & \best{10} & \best{5} & 15\\
    
    selective (with ammuina) & 15 & \best{5} & \best{5} & 35 & 10 & 15 & \best{5} & 20 & \best{5} & 25\\
    
    hierarchical (with ammuina) & 20 & \best{5} & \best{5} & \best{5} & 10 & \best{5} & \best{5} & 15 & \best{5} & 10\\
    
    dynamic (with ammuina) & 40 & \best{5} & 30 & \best{5} & \best{5} & \best{5} & 35 & 15 & \best{5} & 15\\ \midrule
    
    afl++ & 25 & \best{5} & \best{5} & --- & 10 & 20 & \best{5} & 35 & \best{5} & 20\\
    
    nfs & 25 & \best{5} & \best{5} & 20 & 10 & \best{5} & \best{5} & \best{10} & \best{5} & 55\\
    
    \bottomrule
    \end{tabular}}
    \caption{50\% of target coverage.}
    \label{tab:tab:cov_asan_50}
    \end{subtable}
    \begin{subtable}[t]{\linewidth}
    \resizebox{\linewidth}{!}{%
    \begin{tabular}{lcccccccccc}
    \toprule
    \textbf{Poliy} & \textbf{freetype2} & \textbf{guetzli} & \textbf{harfbuzz} & \textbf{lcms} & \textbf{libjpeg} & \textbf{libpng} & \textbf{pcre2} & \textbf{proj4} & \textbf{re2} & \textbf{woff2}\\
    \midrule

    selective & 40 & \best{5} & 10 & --- & 10 & --- & \best{5} & 35 & \best{5} & 40\\
    
    hierarchical & \best{20} & \best{5} & \best{5} & --- & \best{5} & 10 & \best{5} & \best{25} & \best{5} & 55\\
    
    dynamic & 30 & \best{5} & 10 & --- & 15 & \best{5} & 25 & 30 & \best{5} & 15\\
    
    selective (with ammuina) & 45 & \best{5} & 10 & 35 & 10 & --- & \best{5} & 45 & \best{5} & ---\\
    
    hierarchical (with ammuina) & --- & \best{5} & 10 & \best{20} & 10 & 10 & \best{5} & 30 & \best{5} & ---\\
    
    dynamic (with ammuina) & 50 & \best{5} & 35 & 40 & \best{5} & 10 & 50 & --- & \best{5} & \best{10}\\ \midrule
    
    afl++ & 45 & \best{5} & 25 & --- & 25 & 20 & 10 & 35 & \best{5} & ---\\
    
    nfs & 45 & \best{5} & 25 & --- & 10 & 10 & 15 & 50 & \best{5} & ---\\
    
    \bottomrule
    \end{tabular}}
    \caption{75\% of target coverage.}
    \label{tab:tab:cov_asan_75}
    \end{subtable}
    \begin{subtable}[t]{\linewidth}
    \resizebox{\linewidth}{!}{%
    \begin{tabular}{lcccccccccc}
    \toprule
    \textbf{Policy} & \textbf{freetype2} & \textbf{guetzli} & \textbf{harfbuzz} & \textbf{lcms} & \textbf{libjpeg} & \textbf{libpng} & \textbf{pcre2} & \textbf{proj4} & \textbf{re2} & \textbf{woff2}\\
    \midrule

    selective & --- & \best{5} & --- & --- & 25 & --- & 25 & --- & \best{5} & 40\\
    
    hierarchical & \best{30} & \best{5} & \best{15} & --- & 15 & 25 & \best{15} & --- & \best{5} & ---\\
    
    dynamic & 55 & \best{5} & \best{15} & --- & \best{10} & \best{15} & 30 & 55 & \best{5} & \best{20}\\
    
    selective (with ammuina) & --- & \best{5} & 20 & \best{45} & 15 & --- & 20 & --- & \best{5} & ---\\
    
    hierarchical (with ammuina) & --- & \best{5} & \best{15} & --- & 15 & 20 & 25 & --- & \best{5} & 55\\
    
    dynamic (with ammuina) & --- & \best{5} & 40 & --- & 15 & --- & 55 & --- & \best{5} & ---\\ \midrule
    
    afl++ & --- & \best{5} & 45 & --- & 25 & 35 & 45 & \best{35} & \best{5} & ---\\
    
    nfs & --- & \best{5} & 25 & --- & 15 & --- & 25 & --- & \best{5} & ---\\
    
    \bottomrule
    \end{tabular}}
    \caption{90\% of target coverage.}
    \label{tab:tab:cov_asan_90}
    \end{subtable}
    \acmcaption{Time (in minutes) to reach a specific target coverage---ASAN fuzzers.}
    \label{tab:cov_asan}
\end{table}

\begin{table}[t]
    \centering
    \begin{subtable}[t]{\linewidth}
    \resizebox{\linewidth}{!}{%
    \begin{tabular}{lcccccccccc}
    \toprule
    \textbf{Policy} & \textbf{freetype2} & \textbf{guetzli} & \textbf{harfbuzz} & \textbf{lcms} & \textbf{libjpeg} & \textbf{libpng} & \textbf{pcre2} & \textbf{proj4} & \textbf{re2} & \textbf{woff2}\\
    \midrule

    selective & 20 & \best{5} & \best{5} & --- & \best{5} & 30 & \best{5} & 15 & 10 & 40\\
    
    hierarchical & \best{10} & --- & \best{5} & --- & 15 & \best{5} & \best{5} & \best{10} & \best{5} & \best{10}\\
    
    dynamic & 30 & 45 & \best{5} & --- & \best{5} & \best{5} & 25 & \best{10} & \best{5} & 40\\
    
    selective (with ammuina) & 20 & 10 & \best{5} & 35 & 10 & 30 & \best{5} & 20 & 15 & 30\\
    
    hierarchical (with ammuina) & 25 & --- & \best{5} & \best{5} & \best{5} & \best{5} & \best{5} & 15 & 10 & \best{10}\\
    
    dynamic (with ammuina) & 50 & \best{5} & 50 & 10 & \best{5} & 10 & \best{5} & 15 & \best{5} & 60\\ \midrule
    
    afl++ & 25 & 20 & 15 & 60 & 10 & 10 & 10 & 15 & 20 & 35\\
    
    nfs & 25 & 35 & 20 & --- & 10 & 10 & 15 & 15 & 20 & ---\\
    
    \bottomrule
    \end{tabular}}
    \caption{50\% of target coverage.}
    \label{tab:tab:cov_cmplog_50}
    \end{subtable}
    \begin{subtable}[t]{\linewidth}
    \resizebox{\linewidth}{!}{%
    \begin{tabular}{lcccccccccc}
    \toprule
    \textbf{Policy} & \textbf{freetype2} & \textbf{guetzli} & \textbf{harfbuzz} & \textbf{lcms} & \textbf{libjpeg} & \textbf{libpng} & \textbf{pcre2} & \textbf{proj4} & \textbf{re2} & \textbf{woff2}\\
    \midrule

    selective & 45 & \best{5} & 10 & --- & 15 & --- & 15 & 30 & 20 & ---\\
    
    hierarchical & \best{25} & --- & \best{5} & --- & 15 & 15 & \best{10} & \best{15} & \best{5} & ---\\
    
    dynamic & 45 & 45 & 10 & --- & \best{10} & \best{10} & 30 & 25 & 10 & 60\\
    
    selective (with ammuina) & --- & 10 & 10 & 35 & 15 & 60 & 15 & 40 & 20 & ---\\
    
    hierarchical (with ammuina) & --- & --- & 10 & \best{5} & 15 & \best{10} & 15 & 25 & 10 & ---\\
    
    dynamic (with ammuina) & --- & \best{5} & 60 & 10 & \best{10} & --- & \best{10} & 40 & 10 & ---\\ \midrule
    
    afl++ & 45 & 20 & 25 & 60 & 15 & 35 & 25 & 35 & 25 & \best{40}\\
    
    nfs & 45 & 35 & 25 & --- & 15 & --- & 25 & 40 & 25 & ---\\
    
    \bottomrule
    \end{tabular}}
    \caption{75\% of target coverage.}
    \label{tab:tab:cov_cmplog_75}
    \end{subtable}
    \begin{subtable}[t]{\linewidth}
    \resizebox{\linewidth}{!}{%
    \begin{tabular}{lcccccccccc}
    \toprule
    \textbf{Policy} & \textbf{freetype2} & \textbf{guetzli} & \textbf{harfbuzz} & \textbf{lcms} & \textbf{libjpeg} & \textbf{libpng} & \textbf{pcre2} & \textbf{proj4} & \textbf{re2} & \textbf{woff2}\\
    \midrule

    selective & --- & --- & --- & --- & 20 & --- & \best{30} & 60 & 60 & ---\\
    
    hierarchical & \best{30} & --- & \best{15} & --- & 15 & 15 & 40 & --- & 20 & ---\\
    
    dynamic & 60 & --- & \best{15} & --- & \best{10} & \best{10} & 35 & 40 & \best{15} & ---\\
    
    selective (with ammuina) & --- & --- & \best{15} & \best{40} & 15 & 60 & \best{30} & 60 & 20 & ---\\
    
    hierarchical (with ammuina) & --- & --- & \best{15} & --- & 15 & \best{10} & --- & --- & \best{15} & ---\\
    
    dynamic (with ammuina) & --- & \best{15} & --- & --- & 15 & --- & 50 & --- & \best{15} & ---\\ \midrule
    
    afl++ & --- & --- & 45 & --- & 20 & 35 & 45 & \best{35} & 40 & \best{45}\\
    
    nfs & --- & --- & 45 & --- & 15 & --- & --- & --- & 25 & ---\\
    
    \bottomrule
    \end{tabular}}
    \caption{90\% of target coverage.}
    \label{tab:tab:cov_cmplog_90}
    \end{subtable}
    \acmcaption{Time (in minutes) to reach a specific target coverage--- cmplog fuzzers.}
    \label{tab:cov_cmplog}
\end{table}

\begin{table}[t]
    \centering
    \begin{subtable}[t]{\linewidth}
    \resizebox{\linewidth}{!}{%
    \begin{tabular}{lcccccccccc}
    \toprule
    \textbf{Policy} & \textbf{freetype2} & \textbf{guetzli} & \textbf{harfbuzz} & \textbf{lcms} & \textbf{libjpeg} & \textbf{libpng} & \textbf{pcre2} & \textbf{proj4} & \textbf{re2} & \textbf{woff2}\\
    \midrule

    selective & \best{10} & --- & \best{10} & --- & \best{5} & 60 & \best{5} & 15 & \best{5} & 40\\
    
    hierarchical & \best{10} & --- & \best{10} & --- & 15 & \best{5} & \best{5} & \best{10} & \best{5} & \best{10}\\
    
    dynamic & \best{10} & 45 & \best{10} & --- & \best{5} & \best{5} & 10 & \best{10} & \best{5} & 15\\
    
    selective (with ammuina) & 15 & 30 & 15 & 35 & 35 & 25 & \best{5} & 20 & \best{5} & 30\\
    
    hierarchical (with ammuina) & 15 & --- & 10 & \best{5} & \best{5} & 10 & \best{5} & 15 & \best{5} & \best{10}\\
    
    dynamic (with ammuina) & 55 & \best{5} & 10 & \best{5} & \best{5} & 10 & 30 & 15 & 10 & 60\\ \midrule
    
    afl++ & 25 & 20 & 20 & --- & 15 & 10 & \best{5} & 15 & \best{5} & 35\\
    
    nfs & 25 & 35 & 20 & --- & 10 & 10 & \best{5} & 20 & \best{5} & ---\\
    
    \bottomrule
    \end{tabular}}
    \caption{50\% of target coverage.}
    \label{tab:tab:cov_laf_50}
    \end{subtable}
    \begin{subtable}[t]{\linewidth}
    \resizebox{\linewidth}{!}{%
    \begin{tabular}{lcccccccccc}
    \toprule
    \textbf{Policy} & \textbf{freetype2} & \textbf{guetzli} & \textbf{harfbuzz} & \textbf{lcms} & \textbf{libjpeg} & \textbf{libpng} & \textbf{pcre2} & \textbf{proj4} & \textbf{re2} & \textbf{woff2}\\
    \midrule

    selective & 35 & --- & 60 & --- & \best{10} & 60 & \best{5} & 30 & 10 & 55\\
    
    hierarchical & \best{20} & --- & 25 & --- & 15 & \best{5} & \best{5} & --- & 10 & ---\\
    
    dynamic & 30 & 45 & \best{15} & --- & \best{10} & \best{5} & 20 & 20 & \best{5} & \best{30}\\
    
    selective (with ammuina) & 40 & 30 & 35 & \best{35} & 15 & 25 & 10 & 35 & 10 & ---\\
    
    hierarchical (with ammuina) & --- & --- & 25 & --- & 15 & 10 & \best{5} & 25 & 10 & ---\\
    
    dynamic (with ammuina) & --- & \best{15} & 50 & --- & 15 & 10 & 40 & --- & 10 & ---\\ \midrule
    
    afl++ & 45 & 20 & 45 & --- & 15 & 10 & 15 & \best{15} & 25 & 40\\
    
    nfs & 45 & 35 & 25 & --- & \best{10} & 10 & 15 & 40 & 20 & ---\\
    
    \bottomrule
    \end{tabular}}
    \caption{75\% of target coverage.}
    \label{tab:tab:cov_laf_75}
    \end{subtable}
    \begin{subtable}[t]{\linewidth}
    \resizebox{\linewidth}{!}{%
    \begin{tabular}{lcccccccccc}
    \toprule
    \textbf{Policy} & \textbf{freetype2} & \textbf{guetzli} & \textbf{harfbuzz} & \textbf{lcms} & \textbf{libjpeg} & \textbf{libpng} & \textbf{pcre2} & \textbf{proj4} & \textbf{re2} & \textbf{woff2}\\
    \midrule

    selective & 55 & --- & --- & --- & 30 & --- & 20 & --- & 20 & ---\\
    
    hierarchical & \best{30} & --- & --- & --- & 15 & \best{45} & \best{15} & --- & \best{10} & ---\\
    
    dynamic & 55 & --- & \best{25} & --- & \best{10} & --- & 30 & 45 & \best{10} & ---\\
    
    selective (with ammuina) & --- & --- & 50 & \best{45} & 15 & --- & 25 & --- & 25 & ---\\
    
    hierarchical (with ammuina) & --- & --- & 40 & --- & 15 & --- & 45 & --- & \best{10} & ---\\
    
    dynamic (with ammuina) & --- & \best{25} & --- & --- & \best{10} & --- & 40 & --- & \best{10} & ---\\ \midrule
    
    afl++ & 60 & --- & --- & --- & 30 & --- & 25 & \best{45} & 25 & \best{45}\\
    
    nfs & --- & --- & --- & --- & 15 & --- & --- & --- & 25 & ---\\
    
    \bottomrule
    \end{tabular}}
    \caption{90\% of target coverage.}
    \label{tab:tab:cov_laf_90}
    \end{subtable}
    \acmcaption{Time (in minutes) to reach a specific target coverage--- LAF fuzzers.}
    \label{tab:cov_laf}
\end{table}

\begin{table}[t]
    \centering
    \begin{subtable}[t]{\linewidth}
    \resizebox{\linewidth}{!}{%
    \begin{tabular}{lcccccccccc}
    \toprule
    \textbf{Policy} & \textbf{freetype2} & \textbf{guetzli} & \textbf{harfbuzz} & \textbf{lcms} & \textbf{libjpeg} & \textbf{libpng} & \textbf{pcre2} & \textbf{proj4} & \textbf{re2} & \textbf{woff2}\\
    \midrule

    selective & 25 & 30 & 20 & --- & \best{5} & 20 & \best{5} & 15 & 10 & 40\\
    
    hierarchical & \best{15} & 20 & \best{5} & --- & 15 & \best{5} & \best{5} & \best{10} & \best{5} & \best{10}\\
    
    dynamic & 25 & 35 & 10 & --- & \best{5} & \best{5} & 15 & \best{10} & \best{5} & 15\\
    
    selective (with ammuina) & 25 & --- & 15 & 30 & 10 & 20 & \best{5} & 20 & 10 & 25\\
    
    hierarchical (with ammuina) & --- & 20 & 10 & \best{5} & \best{5} & \best{5} & 15 & 15 & 10 & \best{10}\\
    
    dynamic (with ammuina) & 25 & \best{10} & 15 & \best{5} & \best{5} & 10 & \best{5} & 15 & \best{5} & 20\\ \midrule
    
    afl++ & 25 & 15 & 10 & 55 & 10 & 10 & 10 & 15 & \best{5} & 40\\
    
    nfs & 30 & 15 & 10 & --- & 10 & \best{5} & 15 & \best{10} & 10 & 60\\
    
    \bottomrule
    \end{tabular}}
    \caption{50\% of target coverage.}
    \label{tab:tab:cov_other_50}
    \end{subtable}
    \begin{subtable}[t]{\linewidth}
    \resizebox{\linewidth}{!}{%
    \begin{tabular}{lcccccccccc}
    \toprule
    \textbf{Policy} & \textbf{freetype2} & \textbf{guetzli} & \textbf{harfbuzz} & \textbf{lcms} & \textbf{libjpeg} & \textbf{libpng} & \textbf{pcre2} & \textbf{proj4} & \textbf{re2} & \textbf{woff2}\\
    \midrule

    selective & --- & 50 & --- & --- & \best{10} & 35 & 20 & 35 & 20 & 40\\
    
    hierarchical & \best{25} & 40 & 35 & --- & 15 & 10 & \best{10} & \best{30} & \best{10} & ---\\
    
    dynamic & 50 & 45 & \best{15} & --- & \best{10} & \best{5} & 30 & \best{30} & \best{10} & \best{15}\\
    
    selective (with ammuina) & --- & --- & 30 & 35 & 15 & 35 & 20 & 50 & 20 & ---\\
    
    hierarchical (with ammuina) & --- & \best{25} & 25 & \best{5} & \best{10} & 10 & --- & 40 & \best{10} & 55\\
    
    dynamic (with ammuina) & --- & 30 & --- & 10 & \best{10} & 25 & 15 & --- & \best{10} & ---\\ \midrule
    
    afl++ & 45 & \best{25} & --- & 60 & 15 & 15 & 20 & 35 & 15 & 40\\
    
    nfs & 60 & 30 & 40 & --- & \best{10} & 10 & 20 & 45 & 15 & ---\\
    
    \bottomrule
    \end{tabular}}
    \caption{75\% of target coverage.}
    \label{tab:tab:cov_other_75}
    \end{subtable}
    \begin{subtable}[t]{\linewidth}
    \resizebox{\linewidth}{!}{%
    \begin{tabular}{lcccccccccc}
    \toprule
    \textbf{Policy} & \textbf{freetype2} & \textbf{guetzli} & \textbf{harfbuzz} & \textbf{lcms} & \textbf{libjpeg} & \textbf{libpng} & \textbf{pcre2} & \textbf{proj4} & \textbf{re2} & \textbf{woff2}\\
    \midrule

    selective & --- & --- & --- & --- & 20 & --- & 55 & --- & --- & ---\\
    
    hierarchical & \best{35} & 50 & --- & --- & 15 & 25 & 40 & --- & \best{10} & ---\\
    
    dynamic & 60 & --- & \best{15} & --- & \best{10} & \best{15} & \best{35} & 55 & 15 & 55\\
    
    selective (with ammuina) & --- & --- & 50 & \best{35} & 15 & 60 & 45 & --- & 20 & ---\\
    
    hierarchical (with ammuina) & --- & \best{30} & 35 & --- & \best{10} & 20 & --- & --- & \best{10} & ---\\
    
    dynamic (with ammuina) & --- & --- & --- & --- & \best{10} & --- & 50 & --- & \best{10} & ---\\ \midrule
    
    afl++ & --- & 40 & --- & --- & 15 & 35 & \best{35} & \best{35} & 25 & \best{40}\\
    
    nfs & --- & --- & --- & --- & 15 & --- & --- & --- & 25 & ---\\
    
    \bottomrule
    \end{tabular}}
    \caption{90\% of target coverage.}
    \label{tab:tab:cov_other_90}
    \end{subtable}
    \acmcaption{Time (in minutes) to reach a specific target coverage---other fuzzers.}
    \label{tab:cov_other}
\end{table}

Interestingly, some policies provide a program's coverage increase that is more stable over time than others. For example, in Figure~\ref{fig:woff2}, we observe that the dynamic dissemination policy shows a different coverage profile when running with or without ammuina, which is not observed, e.g., in the case of the hierarchical dissemination policy. This behaviour is not observed, for example, in the case of the pcre2 benchmark (see Figure~\ref{fig:pcre2}). The reason of this variability is associated with the fact that such policies may disseminate the input across the various nodes in different orders across different runs. In turn, this change in the order can affect the pace at which the coverage increases over time, because new code paths could be reached later, based on what inputs are processed first. This phenomenon becomes exacerbated (hence, the higher variability) if running the program on a single input takes longer. Indeed, a higher variability has been observed for those programs that require a slightly longer execution time to process a single input.

All the results discussed are summarized in Tables \ref{tab:cov_asan} through \ref{tab:cov_other}. We report the time required to achieve a specific level of coverage. As mentioned, we conducted all fuzzing campaigns over a total period of 6 hours. After this time, we recorded the highest coverage achieved by any of the fuzzers and considered it the ``best coverage'', i.e. the maximum coverage reached by the most effective strategy following an extensive fuzzing process.

As mentioned, our primary goal is to improve fuzzing speed, particularly during the initial phase, so as to provide timely indications of the presence of bugs or vulnerabilities. This is especially relevant when fuzzing is integrated into CI/CD pipelines. Therefore, we have assessed the strategies best suited to accomplish this objective. In the tables, we present three different intermediate target coverages, calculated as 50\%, 75\%, and 90\% of the best coverage for each application we examined. We have sampled the coverage every 5 minutes of execution. We indicate with a ``---'' the configurations for which the desired level of coverage is never reached.

As can be seen from the tables, the MPI-based policies are capable of providing desired results faster, particularly when the considered target coverage is smaller. This is a clear indication that a careful input exchange can benefit the pace at which fuzzing campaigns advance. In some notable cases (e.g., lcms in Table~\ref{tab:tab:cov_cmplog_90}), ammuina can provide a non-negligible benefit over other configurations.

\begin{table}[t]
    \small
    \begin{subtable}[t]{\linewidth}
    \centering
    \begin{tabular}{lccccccc}
    \toprule
    \textbf{Policy} & \textbf{guetzli} & \textbf{harfbuzz} & \textbf{lcms} & \textbf{libpng} & \textbf{pcre2} & \textbf{proj4} & \textbf{re2} \\
    \midrule
    selective & 41 & 46 & 8 & 14 & 922 & \best{299} & 5\\
    
    hierarchical & 52 & 73 & 6 & 12 & 1050 & 215 & 20\\
    
    dynamic & 35 & 54 & 6 & \best{16} & 1091 & \best{299} & 48\\
    
    selective (with ammuina) & 22 & 27 & 58 & \best{16} & 984 & 214 & 7\\
    
    hierarchical (with ammuina) & 34 & \best{85} & 12 & 11 & 870 & 206 & 22\\
    
    dynamic (with ammuina) & \best{64} & 50 & 12 & 14 & 1002 & 222 & 16\\ \midrule
    
    afl++ & 39 & 44 & 7 & \best{16} & \best{1412} & 219 & 34\\
    
    nfs & 61 & 7 & \best{90} & \best{16} & 649 & 181 & \best{139}\\
    
    \bottomrule
    \end{tabular}
    \caption{Maximum number of crashes observed}
    \label{tab:max_crashes}
    \end{subtable}
    
    \begin{subtable}[t]{\linewidth}
    \centering
    \begin{tabular}{lccccccc}
    \toprule
    \textbf{Policy} & \textbf{guetzli} & \textbf{harfbuzz} & \textbf{lcms} & \textbf{libpng} & \textbf{pcre2} & \textbf{proj4} & \textbf{re2} \\
    \midrule
    selective & \best{63577} & 336203 & 2628 & 12166 & 25850 & 35721 & 424159\\
    
    hierarchical & 213668 & \best{318314} & 2826 & 17557 & 31105 & 183129 & 459261\\
    
    dynamic & 330553 & 603351 & 2300 & 15738 & 520300 & 70984 & 805665\\
    
    selective (with ammuina) & 87384 & 842310 & 2509 & 25817 & 24660 & \best{24760} & \best{262321}\\
    
    hierarchical (with ammuina) & 109181 & 517126 & 2910 & 42997 & 31731 & 29944 & 522950\\
    
    dynamic (with ammuina) & 197520 & 1366570 & 6499 & 21547 & 76103 & 105554 & 369158\\ \midrule
    
    afl++ & 110454 & 394404 & 349 & 13124 & \best{16443} & 111116 & 827796\\
    
    nfs & 134257 & 547241 & \best{232} & \best{7363} & 17834 & 109460 & 318916\\
    
    \bottomrule
    \end{tabular}
    \caption{Time to observe the first crash (ms)}
    \label{tab:crash_time}
    \end{subtable}
    \caption{Capability to identify bugs}
    \label{tab:crash}
\end{table}

These results are also confirmed in Table~\ref{tab:crash}, where we report the maximum number of crashes observed when employing the different fuzzing strategies. As can be seen in Table~\ref{tab:max_crashes}, employing our MPI-based policies does not necessarily improve the capability of identifying bugs, although some configurations slightly improve it. This is expected, as we are not dealing with the way inputs are generated or mutated, but only with their dissemination among fuzzing nodes. Notably, in most configurations, our approach can reduce the time to observe the first crash (Table~\ref{tab:crash_time}), as we already discussed above. Interestingly, for applications that are harder to fuzz (e.g., pcre2), the introduction of ammunica can reduce this time by an order of magnitude.

\section{Conclusions and Future Work}

The experimental assessment that we have conducted demonstrates that using MPI-based synchronization can improve distributed fuzzing performance. Lightweight MPI primitives can reduce latency compared to filesystem-based methods, with benefits especially clear when testing ASAN-instrumented binaries.

Among the policies evaluated, the hierarchical approach accelerates initial coverage by enabling efficient intra-cluster input dissemination. Over time, however, the dynamic policy outperforms others by adaptively routing inputs where they are most effective, highlighting the value of real-time resource feedback.

Ammuina mode complements these strategies by periodically triggering broader input exchange if coverage exploration stagnates. Despite the added overhead, it often restores progress, as seen with HarfBuzz and LCMS, where performance improved after ammuina activations.

Combining MPI synchronization with AFL++ presents a promising approach to achieving more scalable fuzzing. These results may have timidly opened a new path between these two (only apparently) distant research areas, encouraging greater collaboration between these domains. In future work, we plan to focus on dynamically selecting the most appropriate policy from the various options we have introduced. Given that different applications can exhibit distinct behaviors, automatically determining the best policy at runtime could further enhance the effectiveness of distributed fuzzing campaigns.

\section*{Acknowledgements}
The authors thank Dr. Andrea Piccione and Dr. Adriano Pimpini for the initial ideas and discussions that have led to this work.

\bibliographystyle{ACM-Reference-Format}
\bibliography{references,pierrefs}


\begin{thebibliography}{24}


\ifx \showCODEN    \undefined \def \showCODEN     #1{\unskip}     \fi
\ifx \showDOI      \undefined \def \showDOI       #1{#1}\fi
\ifx \showISBNx    \undefined \def \showISBNx     #1{\unskip}     \fi
\ifx \showISBNxiii \undefined \def \showISBNxiii  #1{\unskip}     \fi
\ifx \showISSN     \undefined \def \showISSN      #1{\unskip}     \fi
\ifx \showLCCN     \undefined \def \showLCCN      #1{\unskip}     \fi
\ifx \shownote     \undefined \def \shownote      #1{#1}          \fi
\ifx \showarticletitle \undefined \def \showarticletitle #1{#1}   \fi
\ifx \showURL      \undefined \def \showURL       {\relax}        \fi
\providecommand\bibfield[2]{#2}
\providecommand\bibinfo[2]{#2}
\providecommand\natexlab[1]{#1}
\providecommand\showeprint[2][]{arXiv:#2}

\bibitem[Afzal et~al\mbox{.}(2017)]%
        {Afzal2017-ta}
\bibfield{author}{\bibinfo{person}{Asif Afzal}, \bibinfo{person}{Zahid Ansari}, \bibinfo{person}{Ahmed~Rimaz Faizabadi}, {and} \bibinfo{person}{M~K Ramis}.} \bibinfo{year}{2017}\natexlab{}.
\newblock \showarticletitle{Parallelization strategies for computational fluid dynamics software: State of the art review}.
\newblock \bibinfo{journal}{\emph{Archives of Computational Methods in Engineering. State of the Art Reviews}} \bibinfo{volume}{24}, \bibinfo{number}{2} (\bibinfo{date}{April} \bibinfo{year}{2017}), \bibinfo{pages}{337--363}.
\newblock
\showISSN{1134-3060,1886-1784}
\urldef\tempurl%
\url{https://doi.org/10.1007/s11831-016-9165-4}
\showDOI{\tempurl}


\bibitem[Aschermann et~al\mbox{.}(2019)]%
        {Aschermann2019-kr}
\bibfield{author}{\bibinfo{person}{Cornelius Aschermann}, \bibinfo{person}{Sergej Schumilo}, \bibinfo{person}{Tim Blazytko}, \bibinfo{person}{Robert Gawlik}, {and} \bibinfo{person}{Thorsten Holz}.} \bibinfo{year}{2019}\natexlab{}.
\newblock \showarticletitle{{REDQUEEN}: Fuzzing with Input-to-State Correspondence}. In \bibinfo{booktitle}{\emph{Symposium on Network and Distributed System Security (NDSS)}}.
\newblock


\bibitem[Carothers et~al\mbox{.}(2002)]%
        {Carothers2002-ue}
\bibfield{author}{\bibinfo{person}{Christopher Carothers}, \bibinfo{person}{David Bauer}, {and} \bibinfo{person}{Shawn Pearce}.} \bibinfo{year}{2002}\natexlab{}.
\newblock \showarticletitle{{ROSS}: A high-performance, low-memory, modular Time Warp system}.
\newblock \bibinfo{journal}{\emph{Journal of parallel and distributed computing}} \bibinfo{volume}{62}, \bibinfo{number}{11} (\bibinfo{date}{Nov.} \bibinfo{year}{2002}), \bibinfo{pages}{1648--1669}.
\newblock
\showISSN{0743-7315}
\urldef\tempurl%
\url{https://doi.org/10.1016/S0743-7315(02)00004-7}
\showDOI{\tempurl}


\bibitem[Chen et~al\mbox{.}(2018)]%
        {Chen2018-ph}
\bibfield{author}{\bibinfo{person}{Yuanliang Chen}, \bibinfo{person}{Yu Jiang}, \bibinfo{person}{Fuchen Ma}, \bibinfo{person}{Jie Liang}, \bibinfo{person}{Mingzhe Wang}, \bibinfo{person}{Chijin Zhou}, \bibinfo{person}{Xun Jiao}, {and} \bibinfo{person}{Zhuo Su}.} \bibinfo{year}{2018}\natexlab{}.
\newblock \showarticletitle{{EnFuzz}: Ensemble fuzzing with seed synchronization among diverse fuzzers}. In \bibinfo{booktitle}{\emph{Proceedings of the 28th USENIX Security Symposium}} \emph{(\bibinfo{series}{USENIX Security'19})}. \bibinfo{publisher}{Usenix Association}, \bibinfo{address}{Santa Clara, CA, USA}, \bibinfo{pages}{1967--1983}.
\newblock


\bibitem[Christidis(2015)]%
        {Christidis2015-gz}
\bibfield{author}{\bibinfo{person}{Zaphiris Christidis}.} \bibinfo{year}{2015}\natexlab{}.
\newblock \showarticletitle{Performance and scaling of {WRF} on three different parallel supercomputers}.
\newblock In \bibinfo{booktitle}{\emph{Lecture Notes in Computer Science}}. \bibinfo{publisher}{Springer International Publishing}, \bibinfo{address}{Cham}, \bibinfo{pages}{514--528}.
\newblock
\showISBNx{9783319201184,9783319201191}
\showISSN{1611-3349,0302-9743}
\urldef\tempurl%
\url{https://doi.org/10.1007/978-3-319-20119-1\_37}
\showDOI{\tempurl}


\bibitem[Collet({[n.\,d.]})]%
        {ColletUnknown-rt}
\bibfield{author}{\bibinfo{person}{Yann Collet}.} \bibinfo{year}{[n.\,d.]}\natexlab{}.
\newblock \bibinfo{title}{{xxHash}: Extremely Fast Non-cryptographic Hash Algorithm}.
\newblock
\newblock


\bibitem[Fioraldi et~al\mbox{.}(2020)]%
        {Fioraldi2020-pz}
\bibfield{author}{\bibinfo{person}{Andrea Fioraldi}, \bibinfo{person}{D Maier}, \bibinfo{person}{H Eißfeldt}, {and} \bibinfo{person}{Marc Heuse}.} \bibinfo{year}{2020}\natexlab{}.
\newblock \showarticletitle{{AFL++} : Combining incremental steps of fuzzing research}. In \bibinfo{booktitle}{\emph{Proceedings of the 14th USENIX Conference on Offensive Technologies}} \emph{(\bibinfo{series}{WOOT'20})}. \bibinfo{publisher}{USENIX Association}, \bibinfo{address}{Santa Clara, CA, USA}, \bibinfo{pages}{1--12}.
\newblock
\urldef\tempurl%
\url{https://doi.org/10.5555/3488877.3488887}
\showDOI{\tempurl}


\bibitem[Gabriel et~al\mbox{.}(2004)]%
        {Gabriel2004-nr}
\bibfield{author}{\bibinfo{person}{Edgar Gabriel}, \bibinfo{person}{Graham~E Fagg}, \bibinfo{person}{George Bosilca}, \bibinfo{person}{Thara Angskun}, \bibinfo{person}{Jack~J Dongarra}, \bibinfo{person}{Jeffrey~M Squyres}, \bibinfo{person}{Vishal Sahay}, \bibinfo{person}{Prabhanjan Kambadur}, \bibinfo{person}{Brian Barrett}, \bibinfo{person}{Andrew Lumsdaine}, \bibinfo{person}{Ralph~H Castain}, \bibinfo{person}{David~J Daniel}, \bibinfo{person}{Richard~L Graham}, {and} \bibinfo{person}{Timothy~S Woodall}.} \bibinfo{year}{2004}\natexlab{}.
\newblock \showarticletitle{Open {MPI}: Goals, concept, and design of a next generation {MPI} implementation}.
\newblock In \bibinfo{booktitle}{\emph{Recent Advances in Parallel Virtual Machine and Message Passing Interface}}. \bibinfo{publisher}{Springer Berlin Heidelberg}, \bibinfo{address}{Berlin, Heidelberg}, \bibinfo{pages}{97--104}.
\newblock
\showISBNx{9783540231639,9783540302186}
\showISSN{0302-9743,1611-3349}
\urldef\tempurl%
\url{https://doi.org/10.1007/978-3-540-30218-6\_19}
\showDOI{\tempurl}


\bibitem[Gan et~al\mbox{.}(2018)]%
        {Gan2018-tn}
\bibfield{author}{\bibinfo{person}{Shuitao Gan}, \bibinfo{person}{Chao Zhang}, \bibinfo{person}{Xiaojun Qin}, \bibinfo{person}{Xuwen Tu}, \bibinfo{person}{Kang Li}, \bibinfo{person}{Zhongyu Pei}, {and} \bibinfo{person}{Zuoning Chen}.} \bibinfo{year}{2018}\natexlab{}.
\newblock \showarticletitle{{CollAFL}: Path Sensitive Fuzzing}. In \bibinfo{booktitle}{\emph{2018 IEEE Symposium on Security and Privacy (SP)}}. \bibinfo{publisher}{IEEE}, \bibinfo{address}{Piscataway, NJ, USA}, \bibinfo{pages}{679--696}.
\newblock
\showISBNx{9781538643532,9781538643549}
\showISSN{2375-1207}
\urldef\tempurl%
\url{https://doi.org/10.1109/sp.2018.00040}
\showDOI{\tempurl}


\bibitem[Jang et~al\mbox{.}(2022)]%
        {Jang2022-lj}
\bibfield{author}{\bibinfo{person}{Daehee Jang}, \bibinfo{person}{Ammar Askar}, \bibinfo{person}{Insu Yun}, \bibinfo{person}{Stephen Tong}, \bibinfo{person}{Yiqin Cai}, {and} \bibinfo{person}{Taesoo Kim}.} \bibinfo{year}{2022}\natexlab{}.
\newblock \showarticletitle{Fuzzing@home: Distributed fuzzing on untrusted heterogeneous clients}. In \bibinfo{booktitle}{\emph{Proceedings of the 25th International Symposium on Research in Attacks, Intrusions and Defenses}} \emph{(\bibinfo{series}{RAID'22})}. \bibinfo{publisher}{ACM}, \bibinfo{address}{New York, NY, USA}, \bibinfo{pages}{1--16}.
\newblock
\urldef\tempurl%
\url{https://doi.org/10.1145/3545948.3545971}
\showDOI{\tempurl}


\bibitem[Li et~al\mbox{.}(2018b)]%
        {Li2018-np}
\bibfield{author}{\bibinfo{person}{Jun Li}, \bibinfo{person}{Bodong Zhao}, {and} \bibinfo{person}{Chao Zhang}.} \bibinfo{year}{2018}\natexlab{b}.
\newblock \showarticletitle{Fuzzing: a survey}.
\newblock \bibinfo{journal}{\emph{Cybersecurity}} \bibinfo{volume}{1}, \bibinfo{number}{1} (\bibinfo{date}{June} \bibinfo{year}{2018}), \bibinfo{pages}{1--13}.
\newblock
\showISSN{2523-3246,2523-3246}
\urldef\tempurl%
\url{https://doi.org/10.1186/s42400-018-0002-y}
\showDOI{\tempurl}


\bibitem[Li et~al\mbox{.}(2018a)]%
        {Li2018-lk}
\bibfield{author}{\bibinfo{person}{Yang Li}, \bibinfo{person}{Chao Feng}, {and} \bibinfo{person}{Chaojing Tang}.} \bibinfo{year}{2018}\natexlab{a}.
\newblock \showarticletitle{A large-scale parallel fuzzing system}. In \bibinfo{booktitle}{\emph{Proceedings of the 2nd International Conference on Advances in Image Processing}} \emph{(\bibinfo{series}{ICAIP'18})}. \bibinfo{publisher}{ACM}, \bibinfo{address}{New York, NY, USA}, \bibinfo{pages}{194--197}.
\newblock
\showISBNx{9781450364607}
\urldef\tempurl%
\url{https://doi.org/10.1145/3239576.3239615}
\showDOI{\tempurl}


\bibitem[Luo et~al\mbox{.}(2024)]%
        {Luo2024-qj}
\bibfield{author}{\bibinfo{person}{Zhengxiong Luo}, \bibinfo{person}{Junze Yu}, \bibinfo{person}{Qingpeng Du}, \bibinfo{person}{Yanyang Zhao}, \bibinfo{person}{Feifan Wu}, \bibinfo{person}{Heyuan Shi}, \bibinfo{person}{Wanli Chang}, {and} \bibinfo{person}{Yu Jiang}.} \bibinfo{year}{2024}\natexlab{}.
\newblock \showarticletitle{Parallel fuzzing of {IoT} messaging protocols through collaborative packet generation}.
\newblock \bibinfo{journal}{\emph{IEEE transactions on computer-aided design of integrated circuits and systems}} \bibinfo{volume}{43}, \bibinfo{number}{11} (\bibinfo{date}{Nov.} \bibinfo{year}{2024}), \bibinfo{pages}{3431--3442}.
\newblock
\showISSN{0278-0070,1937-4151}
\urldef\tempurl%
\url{https://doi.org/10.1109/tcad.2024.3444705}
\showDOI{\tempurl}


\bibitem[Martin et~al\mbox{.}(2003)]%
        {Martin2003-rh}
\bibfield{author}{\bibinfo{person}{D~E Martin}, \bibinfo{person}{P~A Wilsey}, \bibinfo{person}{R~J Hoekstra}, \bibinfo{person}{E~R Keiter}, \bibinfo{person}{S~A Hutchinson}, \bibinfo{person}{T~V Russo}, {and} \bibinfo{person}{L~J Waters}.} \bibinfo{year}{2003}\natexlab{}.
\newblock \showarticletitle{Redesigning the {WARPED} simulation kernel for analysis and application development}. In \bibinfo{booktitle}{\emph{Proceedings of the 36th Annual Simulation Symposium}} \emph{(\bibinfo{series}{SIMSYM '03})}. \bibinfo{publisher}{IEEE}, \bibinfo{address}{Piscataway, NJ, USA}, \bibinfo{pages}{216--223}.
\newblock
\showISSN{1080-241X}
\urldef\tempurl%
\url{https://doi.org/10.1109/SIMSYM.2003.1192816}
\showDOI{\tempurl}


\bibitem[{Message Passing Interface Forum}(2023)]%
        {Message-Passing-Interface-Forum2023-ou}
\bibfield{author}{\bibinfo{person}{{Message Passing Interface Forum}}.} \bibinfo{year}{2023}\natexlab{}.
\newblock \bibinfo{booktitle}{\emph{{MPI}: A Message-Passing Interface Standard Version 4.1}}.
\newblock


\bibitem[Pellegrini et~al\mbox{.}(2012)]%
        {Pellegrini2012-eo}
\bibfield{author}{\bibinfo{person}{Alessandro Pellegrini}, \bibinfo{person}{Roberto Vitali}, {and} \bibinfo{person}{Francesco Quaglia}.} \bibinfo{year}{2012}\natexlab{}.
\newblock \showarticletitle{The {ROme} {OpTimistic} Simulator: Core Internals and Programming Model}. In \bibinfo{booktitle}{\emph{Proceedings of the 4th International ICST Conference on Simulation Tools and Techniques}} \emph{(\bibinfo{series}{SIMUTOOLS})}. \bibinfo{publisher}{ICST}, \bibinfo{address}{Brussels, Belgium}, \bibinfo{pages}{96--98}.
\newblock
\showISBNx{9781936968008}
\urldef\tempurl%
\url{https://doi.org/10.4108/icst.simutools.2011.245551}
\showDOI{\tempurl}


\bibitem[Pham et~al\mbox{.}(2021)]%
        {Pham2021-zv}
\bibfield{author}{\bibinfo{person}{Van-Thuan Pham}, \bibinfo{person}{Manh-Dung Nguyen}, \bibinfo{person}{Quang-Trung Ta}, \bibinfo{person}{Toby Murray}, {and} \bibinfo{person}{Benjamin I~P Rubinstein}.} \bibinfo{year}{2021}\natexlab{}.
\newblock \showarticletitle{Towards systematic and dynamic task allocation for collaborative parallel fuzzing}. In \bibinfo{booktitle}{\emph{Proceedings of the 36th IEEE/ACM International Conference on Automated Software Engineering}} \emph{(\bibinfo{series}{ASE'21})}. \bibinfo{publisher}{IEEE}, \bibinfo{address}{Piscataway, NJ, USA}, \bibinfo{pages}{1337--1341}.
\newblock
\showISBNx{9781665403375,9781665447843}
\showISSN{2643-1572,1938-4300}
\urldef\tempurl%
\url{https://doi.org/10.1109/ase51524.2021.9678810}
\showDOI{\tempurl}


\bibitem[Ross(2008)]%
        {Ross2008-fb}
\bibfield{author}{\bibinfo{person}{Philip~E Ross}.} \bibinfo{year}{2008}\natexlab{}.
\newblock \showarticletitle{Why {CPU} frequency stalled}.
\newblock \bibinfo{journal}{\emph{IEEE spectrum}} \bibinfo{volume}{45}, \bibinfo{number}{4} (\bibinfo{date}{April} \bibinfo{year}{2008}), \bibinfo{pages}{72--72}.
\newblock
\showISSN{0018-9235,1939-9340}
\urldef\tempurl%
\url{https://doi.org/10.1109/mspec.2008.4476447}
\showDOI{\tempurl}


\bibitem[Serebryany(2017)]%
        {Serebryany2017-nq}
\bibfield{author}{\bibinfo{person}{Kostya Serebryany}.} \bibinfo{year}{2017}\natexlab{}.
\newblock \showarticletitle{{OSS}-Fuzz - Google's continuous fuzzing service for open source software}. In \bibinfo{booktitle}{\emph{Proceedings of the 26th Usenix Security Symposium}} \emph{(\bibinfo{series}{Usenix Security'17})}. \bibinfo{publisher}{USENIX Association}, \bibinfo{address}{Santa Clara, CA, USA}.
\newblock


\bibitem[Song et~al\mbox{.}(2019)]%
        {Song2019-da}
\bibfield{author}{\bibinfo{person}{Congxi Song}, \bibinfo{person}{Xu Zhou}, \bibinfo{person}{Qidi Yin}, \bibinfo{person}{Xinglu He}, \bibinfo{person}{Hangwei Zhang}, {and} \bibinfo{person}{Kai Lu}.} \bibinfo{year}{2019}\natexlab{}.
\newblock \showarticletitle{{P}-fuzz: A parallel grey-box fuzzing framework}.
\newblock \bibinfo{journal}{\emph{Applied sciences (Basel, Switzerland)}} \bibinfo{volume}{9}, \bibinfo{number}{23} (\bibinfo{date}{Nov.} \bibinfo{year}{2019}), \bibinfo{pages}{5100}.
\newblock
\showISSN{2076-3417}
\urldef\tempurl%
\url{https://doi.org/10.3390/app9235100}
\showDOI{\tempurl}


\bibitem[Thompson et~al\mbox{.}(2022)]%
        {Thompson2022-el}
\bibfield{author}{\bibinfo{person}{Aidan~P Thompson}, \bibinfo{person}{H~Metin Aktulga}, \bibinfo{person}{Richard Berger}, \bibinfo{person}{Dan~S Bolintineanu}, \bibinfo{person}{W~Michael Brown}, \bibinfo{person}{Paul~S Crozier}, \bibinfo{person}{Pieter~J in~'t Veld}, \bibinfo{person}{Axel Kohlmeyer}, \bibinfo{person}{Stan~G Moore}, \bibinfo{person}{Trung~Dac Nguyen}, \bibinfo{person}{Ray Shan}, \bibinfo{person}{Mark~J Stevens}, \bibinfo{person}{Julien Tranchida}, \bibinfo{person}{Christian Trott}, {and} \bibinfo{person}{Steven~J Plimpton}.} \bibinfo{year}{2022}\natexlab{}.
\newblock \showarticletitle{{LAMMPS} - a flexible simulation tool for particle-based materials modeling at the atomic, meso, and continuum scales}.
\newblock \bibinfo{journal}{\emph{Computer physics communications}} \bibinfo{volume}{271}, \bibinfo{number}{108171} (\bibinfo{date}{Feb.} \bibinfo{year}{2022}), \bibinfo{pages}{108171}.
\newblock
\showISSN{0010-4655,1879-2944}
\urldef\tempurl%
\url{https://doi.org/10.1016/j.cpc.2021.108171}
\showDOI{\tempurl}


\bibitem[Van Der~Spoel et~al\mbox{.}(2005)]%
        {Van-Der-Spoel2005-tk}
\bibfield{author}{\bibinfo{person}{David Van Der~Spoel}, \bibinfo{person}{Erik Lindahl}, \bibinfo{person}{Berk Hess}, \bibinfo{person}{Gerrit Groenhof}, \bibinfo{person}{Alan~E Mark}, {and} \bibinfo{person}{Herman J~C Berendsen}.} \bibinfo{year}{2005}\natexlab{}.
\newblock \showarticletitle{{GROMACS}: fast, flexible, and free}.
\newblock \bibinfo{journal}{\emph{Journal of computational chemistry}} \bibinfo{volume}{26}, \bibinfo{number}{16} (\bibinfo{date}{Dec.} \bibinfo{year}{2005}), \bibinfo{pages}{1701--1718}.
\newblock
\showISSN{0192-8651,1096-987X}
\urldef\tempurl%
\url{https://doi.org/10.1002/jcc.20291}
\showDOI{\tempurl}


\bibitem[Wang et~al\mbox{.}(2021)]%
        {Wang2021-zj}
\bibfield{author}{\bibinfo{person}{Yifan Wang}, \bibinfo{person}{Yuchen Zhang}, \bibinfo{person}{Chenbin Pang}, \bibinfo{person}{Peng Li}, \bibinfo{person}{Nikolaos Triandopoulos}, {and} \bibinfo{person}{Jun Xu}.} \bibinfo{year}{2021}\natexlab{}.
\newblock \showarticletitle{Facilitating parallel fuzzing with mutually-exclusive task distribution}.
\newblock In \bibinfo{booktitle}{\emph{Security and Privacy in Communication Networks}}, \bibfield{editor}{\bibinfo{person}{Joaquin Garcia-Alfaro}, \bibinfo{person}{Shujun Li}, \bibinfo{person}{Radha Poovendran}, \bibinfo{person}{Hervé Debar}, {and} \bibinfo{person}{Moti Yung}} (Eds.). \bibinfo{publisher}{Springer International Publishing}, \bibinfo{address}{Cham, Switzerland}, \bibinfo{pages}{185--206}.
\newblock
\showISSN{1867-8211,1867-822X}
\urldef\tempurl%
\url{https://doi.org/10.1007/978-3-030-90022-9\_10}
\showDOI{\tempurl}


\bibitem[Zhou et~al\mbox{.}(2023)]%
        {Zhou2023-ps}
\bibfield{author}{\bibinfo{person}{Xu Zhou}, \bibinfo{person}{Pengfei Wang}, \bibinfo{person}{Chenyifan Liu}, \bibinfo{person}{Tai Yue}, \bibinfo{person}{Yingying Liu}, \bibinfo{person}{Congxi Song}, \bibinfo{person}{Kai Lu}, \bibinfo{person}{Qidi Yin}, {and} \bibinfo{person}{Xu Han}.} \bibinfo{year}{2023}\natexlab{}.
\newblock \showarticletitle{{UltraFuzz}: Towards resource-saving in distributed fuzzing}.
\newblock \bibinfo{journal}{\emph{IEEE transactions on software engineering}} \bibinfo{volume}{49}, \bibinfo{number}{4} (\bibinfo{date}{April} \bibinfo{year}{2023}), \bibinfo{pages}{2394--2412}.
\newblock
\showISSN{0098-5589,1939-3520}
\urldef\tempurl%
\url{https://doi.org/10.1109/tse.2022.3219520}
\showDOI{\tempurl}


\end{thebibliography}

\end{document}